\documentclass[preprint,  aoas]{imsart}

\usepackage{float}
\usepackage{amsmath}
\usepackage{amsfonts}
\usepackage{amssymb}
\usepackage{amsthm}
\usepackage{geometry}
\usepackage{graphicx}
\usepackage{hyperref}
\usepackage{icomma}
\usepackage{latexsym}
\usepackage{textcomp}
\usepackage[all]{xy}
\usepackage{cancel}
\usepackage{color}
\usepackage{dsfont}
\usepackage{subfig}
\usepackage{caption}
\usepackage{dsfont}
\usepackage{extarrows}
\usepackage{shorttoc}
\usepackage{listings}
\usepackage[utf8]{inputenc}
\usepackage{lmodern}
\usepackage{xcolor}
\usepackage{moreverb}

\usepackage[section]{placeins}

\usepackage{breakcites}

\usepackage{booktabs}
\usepackage{graphicx}

\RequirePackage[OT1]{fontenc}
\RequirePackage{amsthm,amsmath,natbib}
\RequirePackage{hypernat}

\usepackage{graphicx,color}

\usepackage{amssymb}

\usepackage{multirow}

\def\be{\begin{equation}}
\def\ee{\end{equation}}
\def\ba{\begin{array}}
\def\ea{\end{array}}

\def\bi{\begin{itemize}}
\def\ei{\end{itemize}}

\usepackage[T1]{fontenc} 
\usepackage[utf8]{inputenc}

\begin{document}
\begin{frontmatter}
\title{
A general theory for preferential sampling in environmental networks.
}
\runtitle{Adjusting for preferential sampling in spatio-temporal analyses}

\begin{aug}
\author{\fnms{Joe} \snm{Watson$^{1}$}\thanksref{m1}\ead[label=e1]{joe.watson@stat.ubc.ca}}
\and 
\author{\fnms{James V} \snm{Zidek$^{1}$}\thanksref{m1}\ead[label=e2]{jim@stat.ubc.ca}}
\and
\author{\fnms{Gavin} \snm{Shaddick}\thanksref{m2}\ead[label=e3]{G.Shaddick@exeter.ac.uk}}
\affiliation{University of British Columbia\thanksmark{m1} and University of British Columbia\thanksmark{m2} 
and University of Exeter\thanksmark{m3} }
\thankstext{m1}{The research reported in this article was partially supported by the Natural Science and Engineering Research Council of Canada.}
\runauthor{Watson and Zidek and Shaddick}
\address{Department of Statistics\\
University of British Columbia\\
2207 Main Mall
Vancouver, BC\\
V6T 1Z4, Canada\\
\printead{e1}\\}

\address{Department of Statistics\\
University of British Columbia\\
2207 Main Mall
Vancouver, BC\\
V6T 1Z4, Canada\\
\printead{e2}\\}

\address{Department of Mathematical Sciences\\
University of Exeter\\
Exeter, Devon \\
EX4 4SB UK\\
\printead{e3}\\
}

\end{aug}

\begin{abstract}

This paper presents a general model framework for detecting the preferential sampling of environmental monitors recording an environmental process across space and/or time. 
This is achieved by considering the joint distribution of an environmental process with a site--selection process that considers where and when sites are placed to measure the process. The environmental process may be spatial, temporal or spatio--temporal in nature. By sharing random effects between the two processes, the joint model is able to establish whether site placement was stochastically dependent of the environmental process under study. Furthermore, if stochastic dependence is identified between the two processes, then inferences about the probability distribution of the spatio--temporal process will change, as will predictions made of the process across space and time. The embedding into a spatio--temporal framework also allows for the modelling of the dynamic site---selection process itself. Real--world factors affecting both the size and location of the network can be easily modelled and quantified. Depending upon the choice of population of locations to consider for selection across space and time under the site--selection process, different insights about the precise nature of preferential sampling can be obtained. The general framework developed in the paper is designed to be easily and quickly fit using the R-INLA package. We apply this framework to a case study involving particulate air pollution over the UK where a major reduction in the size of a monitoring network through time occurred. It is demonstrated that a significant response--biased reduction in the air quality monitoring network occurred, namely the relocation of monitoring sites to locations with the highest pollution levels, and the routine removal of sites at locations with the lowest. We also show that the network was consistently unrepresentative of the levels of particulate matter seen across much of GB throughout the operating life of the network. Finally we show that this may have led to a severe over-reporting of the population--average exposure levels experienced across GB. This could have great impacts on estimates of the health effects of black smoke levels.              

\end{abstract}


\begin{keyword}[class=AMS]
\kwd[Primary ]{62P12}
\kwd[; secondary ]{62D99}
\end{keyword}

\begin{keyword}
\kwd{preferential sampling}
\kwd{INLA}
\kwd{random fields}
\kwd{mobile monitors}
\kwd{health effects}
\kwd{air pollution}
\kwd{Big Data}
\end{keyword}

\end{frontmatter}

\section{Introduction}\label{sect:intro}\label{sect:introduction}
This paper concerns preferential sampling (PS), where the locations of 
sites selected to monitor a spatio--temporal environmental process  
$Z_{st},~s\in {\cal S},~t\in {\cal T}$, depend stochastically on the process they are measuring. Thus PS is a special case of response--biased sampling. The space--time point is defined $(\textbf{s},t)\in {\cal S} \times{T}$, with $\cal S$ denoting the spatial domain of interest and $\cal T$ the temporal domain. Purely spatial processes (i.e. when $|T| = 1$), and purely temporal processes (i.e. when $\cal S$ is ignored) are two special cases. 

 Spatial sampling network designers must specify a set of time points $T \subset {\cal T}$ at which to observe $Z$  and at each time $t\in {T}$, a  finite subset of sites $S_t\subset {\cal S}$ at which to do so. Generally the temporal domain ${\cal T}$ would be a finite set as for practical reasons $Z$ must be a time--averaged quantity. 
%
%
 The designer may select the network sites in a preferential way
to meet specified objectives 
 \citep{schumacher1993using}, although attaining those objectives may 
 present its own challenges 
\citep{chang2007designing}).   
Moreover, the suitability of the network for achieving its initial objectives may decline over 
time as in the case of the air quality monitoring network 
for Metro Vancouver
\citep{ainslie_application_2009}.
In some cases, the objectives may not be well prescribed in which case  
evidence suggests that in these cases administrators may select monitoring sites 
preferentially \citep{shaddick2014case}.  
Finally, the data provided by networks for one purpose may be  used for another purpose and this may cause problems. For example,  
urban air pollution monitoring sites provide the information
needed to detect noncompliance with air quality
standards \citep{ozone05,loperfido2008network}.    
However, these measured values of $Z$ would tend to
overestimate the overall levels of the air pollutant throughout $\cal S$ 
and thus render the data unsuitable for 
assessing the impacts of $Z$ on human health and
welfare. 
 In such cases networks well designed for one 
purpose may be seen as preferentially sampled when the data
they yield are used for another purpose.

A variety of approaches can be taken for modelling PS and mitigating its effects in a spatio--temporal process framework. The choice of framework depends on contexts and purposes. Subsection \ref{subsec:background} reviews some of these approaches along with their associated references.   Two different situations are encountered. In what might be called the retrospective approach all the process data are available for use in assessing and mitigating the impact of PS at any given time $t\leq \text{max} \left(T\right)$. Such impacts could, for example, distort estimates of model parameters, spatial predictions, temporal forecasts, trends, and risk assessments. A special case is  
where $|T|$ $= 1$ and $Z_{sT}, s\in {\cal S}$ is a random spatial field.  Since data are not collected over time, strong assumptions must be made about the preferential sampling process that yields the network of sites.  The data cannot be used to build an emulator of the actual selection process itself, since the requisite data are not yet available when the spatial sites are selected. But it might be assumed that the future latent data does reflect the past during the period under which the network was designed.   

In the prospective case, the selection of network sites at time $t\in T$ may be based on process observations up to and including time $t-1$.  In this case, the propensity to preferentially select sites at time $t$ can be estimated without benefit of having the data for time $t$.  The temporal model can then be sequentially updated at time $t+1$ and the process model could adapt quickly to abrupt changes rather than projecting long term trends.  

We develop a general modelling framework for the retrospective case, that enables a researcher to determine if the locations of the monitoring sites that form an operational network have been selected preferentially through time (i.e. if response--biased selection occurred). Furthermore, unlike with the spatial--only data, our framework applied to spatio--temporal data allows for a site--selection process emulator to be developed. The population of all site locations considered for selection at any time $t \in T$ is defined as $\cal P \subset \cal S$. $\cal P$ must be specified a--priori, as the model framework does not consider locations outside of the fixed (pre--specified) population $\cal P$ in the site--selection process. But within that framework both static and mobile monitoring networks are admitted. Importantly, depending on the choice of population $\cal P$, different insights into the nature of PS can be explored.

Defining the population of sites considered for selection throughout $(\cal S \times \cal T)$ has been an issue of fundamental importance for all previous work on PS. This is especially true for the model framework introduced in this paper. Depending on the choice of population, different insights into the nature of PS can be obtained and spatial predictions may change dramatically. We consider two populations in this paper, however more can be thought of and implemented to suit the needs and knowledge of the researchers. In one case that population is considered to consist of all sites that have been deemed worthy of being monitored at some times $t \in T$. We refer to these as the observed sites. In the other case, pseudo--sites are also included uniformly throughout $\cal S$.  These have never been monitored but are considered important for characterizing the field itself and for investigating the impacts of PS. 
The name pseudo--sites follows from presence-only applications in statistical ecology, where such sites are often referred to as pseudo zeros \citep{warton2010poisson,fithian2013finite}. We opt for the name pseudo--sites to distinguish these locations from the traditional `data-locations' and `prediction-locations' terminology used in classical geostatistics. This is because in many applications, not all prediction locations can also be psuedo site locations. For example there may be regions in $A \subset \cal S$ that we wish to predict the field across, yet know with certainty that a site could not have been considered for selection for reasons unrelated to the process being measured. This could be due to the presence of a physical barrier (e.g. a mountain range) or a political barrier (e.g. a militarised zone) making the placement of a monitoring site impossible.  Note that in all cases our population of sites $\cal P$ is finite. This is in contrast to the spatial continuum assumed by point process models, although parallels between the methodologies exist and are discussed at length in this paper. 

A Bayesian model is introduced for the joint distribution  of the  
response vector $(Y_{st}, R_{st})$. 
$R_{st}$ is a binary response for the site--selection process, which is $0$ or $1$ according 
to whether or not a monitoring site is absent or present at the 
space--time point $(\textbf{s},t)\in {\cal P} \times{T}$, with $\cal P \subset \cal S$ a fixed population of site locations under consideration. The resulting
model when fitted, identifies the effects of PS if any, on inferences about the population mean of the process underlying $Y$. For brevity, we denote this population's mean by `$\cal P$--mean'. By sharing random effects across the two processes, the stochastic dependence (if any) between $Y_{\textbf{s},t}$ and $R_{\textbf{s},t}$ can be quantified, and subsequently the model can adjust the space--time predictions according to the nature of PS detected.

Moreover it yields an emulator of the dynamic preferential site--selection process as the operational monitoring network (denoted by $S_t$) evolves over time. The factors affecting the initial site placements can be allowed to differ from those affecting the retention of existing sites in the network. The dynamic model allows for an assessment of the degree to which preferentiality is determined not just by stochastic processes underlying $Y$, but by other factors that might include for example the administrative processes involved in the establishment of a monitoring site. Two examples considered in this paper are political affinity for environmental monitoring and budgetary constraints in an attempt to emulate the site--selection process, although more can be hypothesised and included. A key result described in the paper is the ability to use the R-INLA software package with the SPDE approach \cite{rue2009approximate,lindgren2011explicit,rue2017bayesian} to fit the joint distributions proposed in our framework. This ensures inference remains feasible, even for space--time applications with many thousands of pseudo--site locations.

Finally, we fit our model framework to a real case study: a large scale air pollution monitoring network in the UK that monitored black smoke (BS hereafter) levels for more than fifty years. This provides an ideal data example for our model since the network underwent a constant, dramatic re--design through time and furthermore, the locations of the observed sites appear to largely under-represent rural regions of Great Britain (GB hereafter). We consider two populations $\cal P$ of sites. First, we consider ${\cal {P}}_1$ to be the locations at which a site was operational at some $t \in T$ (i.e. observed sites only). Here, we ultimately wish to see the effects of PS, if any, on estimates of the ${\cal {P}}_1$--mean, as well as investigate if the network evolved preferentially. Our second population ${\cal {P}}_2$ includes thousands of uniformly located (`pseudo') sites placed at a density of approximately 5km throughout GB. Since we uniformly cover GB, from this population we are able to assess if the observed sites were preferentially placed within GB (i.e. $\cal S$), and then preferentially retained in the network. We can then evaluate the effects of PS on the ${\cal {P}}_2$--mean (i.e. the average across GB). These two choices of population help to address two distinct questions.  

\section{Modelling frameworks}\label{sec:frameworks}
This section describes a very general framework in which PS can be explored depending on the purpose of that exploration.  It begins in Subsection \ref{subsec:background} with a review of some existing theory.

\subsection{Review of related work}\label{subsec:background}
Most work on PS is set in the geostatistical
framework where $T$ consists of a single time point so for expository 
simplicity we temporarily drop the subscript $t$ in this context. In geostatistics PS has a long history.  For example \citet{isaaks1988spatial} describes 
the deleterious impact to variogram estimates when ` the data locations... are preferentially located in high-- or low--valued areas'', in particular because the ``preferentially clustered data'' can lead to a ``destructuring'' of the variogram. In fact this concern about clustered data goes back to  \citet{switzer1977estimation}.   \citet{olea2007declustering} reviews the history of PS, in particular with respect to the clustering due to it. However interest in this topic has spread to a variety of subject areas (see for example \citet{botta2007statistical,michalcova2011bias}).

Interest in the statistical science community seems to have been sparked by the paper of \citet{diggle2010geostatistical} (hereafter DMS). DMS defines the PS of  a space--time field succinctly as the property $[Z, S] \neq [Z][S] $. Here $Z$ denotes the spatial field and $S$ the locations. The square bracket notation can be read as the ``probability distribution of''.   DMS notes that when sampling is non--preferential, $S$ can be regarded as fixed; inferences about $Z$ and its distribution can then be based on conditional distributions given $S$. The authors also note that non--PS differs from ``uniform sampling'' when for a given sample size, every possible realization of $S$ is equally likely. DMS assumes that conditional on $S$ and the  Gaussian process $Z_s,~s\in S$, the measured values of  $Z$ denoted by $Y$ are mutually independent Gaussian random variables with mean $\mu +Z_s$. At the same time, conditional on $Z$, $S$ is assumed to be 
an inhomogeneous Poisson point process with intensity function $\lambda(s)= \exp{\{\alpha + \beta Z_s\}},~s\in {\cal S}$. The parameter $\beta$ represents the degree of PS -- with $\beta~ >~ 0$, implying large values of $Z_s$ are associated with an increased chance of inclusion of a sample in a local neighbourhood around $s$ in $S$. As noted by Professor Dawid in his discussion of DMS, this model cannot represent the real site selection process since the network designers would not know anything about $Z$ until the sites had been established and their measured values were available. Thus this model cannot be viewed as a site--selection emulator since perfect knowledge surrounding $Z$ prior to measurement cannot be assumed.  Nevertheless in a post--hoc analysis of those data, the PS model can be fitted and so capture the impact of the real selection process on inferences made about $Z$ and its probability distribution.  

The inhomogeneous Poisson process model was used subsequent to the publication of DMS by other 
investigators in a similar way but in a fully Bayesian model for inference.
More specifically \citet{gelfand2012effect} replaces $\alpha + \beta Z_s$ in DMS's intensity function by (in our notation) $\alpha + \alpha_1^T {\bf X}_s$ where ${\bf X}$ denotes a vector 
of observable covariates. This change makes the model more like a 
possible model for the real process. Note that without the inclusion of the process $Z_s$ inside the linear predictor of the Poisson process model, they assume a missing--at--random missingness mechanism, with no further dependence existing between the site locations and the underlying process $Z_s$ when conditioned on the included covariates ${\bf X}_s$. Thus this would no longer be considered PS by our earlier definitions. \citet{pati2011bayesian} also includes 
the covariate vector and replaces $\alpha + \beta Z_s$ by 
$\alpha + \alpha_1^T {\bf X}_s +  \beta \xi_s$ so that the effect of 
the observable covariates is incorporated in the PS model. The  $\{\xi_s\}$
are referred to as a ``residual process'' and  so unlike DMS, these authors
are not making PS depend directly on the process $Z$. A second 
residual process $\eta$ is added to the measurement model so conditional on $\xi$, $\eta$, $X$ and $S$ the $\{Y_s\}$ are assumed to be independently distributed with mean $  \mu  + \alpha_1^T {\bf X}_s + \beta \xi_s + \beta_1 \eta_{s}$.  Thus it would seem that in effect that the process model is being represented by $Z_s =  \alpha_1^T {\bf X}_s + \beta \xi_s + \beta_1 \eta_{s}$ while the potential PS derives from only a subcomponent of that process.  

The need to include covariates (predictors) is well recognized in DMS and its ensuing discussions, so \citet{gelfand2012effect} and \citet{pati2011bayesian} are welcome additions to the geostatistical literature on PS. But  none of these models include as we do in this paper, residual terms
that represent the ill--defined administrative and other processes involved in 
actual site selection. These terms are not subcomponents of $Z$ and yet the case study presented in this paper suggests that these residuals play a significant role in PS. Additional work has shown that a failure to properly account for these effects can lead to the overestimation of the magnitude of PS present \citep{Watson3}. Furthermore the point process model on which the above models are based will not be suitable in all applications such as that in \citet{conn2017confronting} about  mapping species abundance in ecology.  That paper presents a general theory for PS where ${\cal S}$ consists of a finite set of points and the response distributions are non--Gaussian to include such things as count data. 



\subsection{A general retrospective modeling framework}\label{subsec:generalframework}
In this section we introduce the general model framework and its purpose, before implementing it on a real case study in Section 4. First, we carefully define the population of locations $\textbf{s} \in \cal P \subset \cal S$ to consider for selection at some or all $t \in T$. The size and placement of this population may substantially affect the resulting inference. In many cases, either the precise locations of all sites under consideration at each $t \in T$ will be known, or there will be a clearly defined population of locations at which interest lies in estimating the space--time field and/or its corresponding population summary statistics. This case is Population 1 (${\cal{P}}_1$) considered in our later application. For the second population (${\cal{P}}_2$) used in our later analysis, we consider all possible points $\textbf{s} \in \cal S$ to be the population. 

Computational considerations lead us, for Population 2, to approximate this by the placement of pseudo--sites in a high density regular grid, thus placing a psuedo site approximately every 5km in $\cal S$. This is similar in flavour to the discretized computational lattice used in the log Gaussian Cox Process (LGCP hereafter) approach by DMS \citep{diggle2010geostatistical}. In fact, as the density of pseudo--sites under consideration in $\cal S$ increases, the resulting logistic regression likelihood converges towards a (scaled) Poisson point process likelihood. Parameter estimates and their standard errors converge to those from the Poisson point process too. However, the accuracy of this approximation depends on the density and placement of the pseudo--sites \citep{warton2010poisson,fithian2013finite}. We discuss this in depth later. The LGCP idea has also been considered further, but the need to explicitly add a third likelihood to the joint model to capture the retention process in spatio--temporal applications may make this approach less desirable in some scenarios \citep{Watson3}.  

Note that the space--time field represented as $Z_{i,t}$ in previous work, is represented in our model framework as a sum of latent random effects. This is done to allow the site--selection process to have independent stochastic dependencies with each of the components making up the space-time field. We let $\cal P$ denote the set of site locations in the population and define $M$ to be the number of sites (i.e. $M = |\cal P| $). Note the interpretation of the $\cal P$--mean differs substantially across these populations. The ${\cal{P}}_1$--mean can be interpreted as the network average, whilst the ${\cal{P}}_2$--mean can be interpreted as the GB--average (the mean of the space--time field across GB).

We let $Y_{i}(t)$ denote a spatio--temporal observation process (continuous, count, etc.) at site $i$, that is at location $\textbf{s}_i \in \cal P \subset S$, at time $t \in T$. We let $R_{i}(t)$ denote the random selection indicator for site $\textbf{s}_i \in P$ at time $t$, with 1 meaning the site was operational at this time. We let $t_1,...,t_N$ denote the (finite) $N$ observation times, and let $r_{i,j} \in \{0,1\}$ denote the realisation of $R_{i}(t_j)$ for site $\textbf{s}_i \in P$ at time $t_j$, $i \in \{1,...,M\}, j \in \{1,...,N\}$. The subscript $j$ will act as a pointer to the desired time. Then our general model framework can be written as follows:

\begin{align}
\left( Y_{i,j} | R_{i,j} = 1 \right) &\backsim f_Y(g(\mu_{i,j}), \boldsymbol{\theta}_Y), \hspace{0.2cm} f_Y \backsim \textnormal{density} \nonumber \\
g(\mu_{i,j}) = \eta_{i,j} &= \textbf{x}^T_{i,j} \boldsymbol{\gamma} + \sum_{k = 1}^{q_1} u_{i,j,k} \beta_{k}(\textbf{s}_i, t_j) \hspace{0.2cm} \nonumber \\
R_{i,j} &\backsim \textnormal{Bernoulli} \left(p_{i,j}\right) \nonumber \\
h(p_{i,j}) = \nu_{i,j} &= \textbf{v}^T_{i,j} \boldsymbol{\alpha} + \sum_{l = 1}^{q_2} d_{l} \sum_{k = 1}^{q_1} w_{i,j,l,k} \beta_{k}(\textbf{s}_i, \phi_{i,l,k}\left(t_j\right))  + \sum_{m = 1}^{q_3} w_{i,j,m}^{\star} \beta^{\star}_{m}(\textbf{s}_i, t_j) \hspace{0.2cm} \nonumber \\
\beta_{k}(\textbf{s}_i, t_j) &\backsim \textnormal{(possibly shared) latent effect with parameters}\hspace{0.15cm} \boldsymbol{\theta_k}  \hspace{0.3cm} k \in \{1,..,q_1\} \nonumber \\
\beta^{\star}_{m}(\textbf{s}_i, t_j)  &\backsim \textnormal{site--selection only latent effect with parameters} \hspace{0.15cm} \boldsymbol{\theta}^{\star}_m  \hspace{0.3cm} m \in \{1,..,q_3\} \nonumber \\
\Theta &= \left(\boldsymbol{\theta_Y}, \boldsymbol{\alpha}, \boldsymbol{\gamma}, \textbf{d}, \boldsymbol{\theta_1}, ... , \boldsymbol{\theta_{q_1}}, \boldsymbol{\theta}^{\star}_1, ... , \boldsymbol{\theta}^{\star}_{q_3} \right) \backsim \textnormal{Priors} \nonumber \\
\textbf{x}_{i,j} \in \mathds{R}^{p_1}, \hspace{0.2cm}
\textbf{u}_{i,j} &\in \mathds{R}^{q_1}, \hspace{0.2cm}\textbf{v}_{i,j} \in \mathds{R}^{p_2}, \hspace{0.2cm}  
\textbf{W}_{i,j} \in \mathds{R}^{q_2 \times q_1}, \hspace{0.2cm} \textbf{w}_{i,j}^{\star T} \in \mathds{R}^{q_3} \nonumber
\end{align}
\\

The above framework is set up to allow for a large degree of modelling flexibility for spatial, temporal and spatio--temporal applications. 
Note that the two functions $g$ and $h$ are known as link functions. These relate the expected value of the response to the linear predictor. Popular choices of $h$ for the Bernoulli likelihood are the logit, complementary log-log and probit functions. In our later analysis, we will generate our zeros (or pseudo--sites) with an approximately constant intensity across $\cal S$. Thus in our case the logit link is the suitable choice for link function since it exploits a natural connection between the conditional logistic regression and the loglinear Poisson point process model we are approximating when we condition on the total count \citep{baddeley2015spatial}.

We now dissect the model term--by--term. Firstly, consider the observation process $Y$. We allow for any distribution to be chosen as the likelihood for the observation process. This allows a range of different data types (e.g. continuous, count, etc.,) to be modelled, including those that exhibit a range of features such as skewness, heavy tails and/or over-dispersion. 
In the linear predictor $\eta_{i,j}$, we may include a linear combination of fixed covariates $\textbf{x}_{i,j}$ with a linear combination of $q_1$ latent effects $\beta_k(\textbf{s}_i,t_j)$. These $q_1$ random effects can include any combination of spatially--correlated processes (such as Gaussian [Markov] random fields), temporally correlated processes (such as autoregressive terms), spatio--temporal processes and IID random effects. Note that we include the additional fixed covariates $\textbf{u}_{i,j}$ to allow for spatially--varying coefficient models, as well as both random slopes and/or scaled random effects to be included. The flexibility here allows for areal data to be modelled too, simply by changing the definition of $\textbf{s}_i$ from being a point to representing a well-defined area. 

Next, we consider the site--selection process $R_{i,j}$. As before, in the linear predictor $\nu_{i,j}$, we may include a linear combination of fixed covariates $\textbf{v}_{i,j}$ with a linear combination of latent effects. This time however, the latent effects appearing in the observation process $Y_{i,j}$ are allowed to exist in the linear predictor of the selection process $R_{i,t}$. It is this feature that allows for stochastic dependence to exist between the two processes and hence enables us to investigate whether we have a missing--not--at--random mechanism. Note that the matrix $\textbf{W}_{i,j}$ is fixed beforehand, and allows for $q_2$ linear combinations (possibly scaled by covariates) of the latent effects from the $Y_{i,j}$ process to be copied across. The parameter vector $\textbf{d}$ determines the degree to which each shared latent effect (or combination of) affects the $R$ process and therefore measures the magnitude and direction of stochastic dependence between the two models term--by--term. We denote this term by $\textbf{d}$ in recognition of the landmark paper by \citet{diggle2010geostatistical}. Finally, as seen in \citet{pati2011bayesian}, we allow $q_3$ latent effects, independent from the $Y_{i,j}$ process to exist in the linear predictor. This allows us to extract as many sources of variation from the site--selection process as possible, reducing the risk of over--estimating the magnitude of the $d_l$ terms, and thus the stochastic dependence between the two processes.

For added flexibility we allow temporal lags in the stochastic dependence. This allows the site--selection process to depend upon the realised values of the latent effects at any arbitrary time in the past, present or future. Thus this framework allows for both proactive and reactive site--selection to occur. For example, if for a pollution monitoring network, site--selection were desired near immediate sources of pollution (say for exceedance detection), then we may view as reasonable, a model that allows for a dependence between the latent field at the previous time step as a site--selection emulator. In this case, we would select as the temporal lag function, $\phi_{i,l,k}\left(t_j\right) = t_{j - 1}$. We define this to be reactive selection, where placement depends only on past realisations of the space--time field. Say instead, site placements were desired near areas forecast to increase in industrialisation (and hence pollution emission). Then a model allowing for dependence with future values of the latent process may be suitable. To achieve this we would select $\phi_{i,l,k}\left(t_j\right) > t_j$. We define this to be proactive site selection. Models with mixtures of reactive and proactive site selection could also be admitted and fit under this framework since a unique temporal lag function $\phi_{i,l,k}\left(t\right)$ is allowed for each latent effect shared between the linear predictors. 

Also of interest is the possibility of setting $w_{i,j,l,m} = 0$ for some values of the subscripts to allow for the directions of preferentiality to change through time. For example, the initial placement of the sites might be made in a positively (or negatively) preferential manner but over time the network might be redesigned so that sites were later placed to reduce the bias. To capture this, it would make sense to have a separate PS parameter $d$ estimated for time $t = 1$ and for times $t > 1$ to capture the changing directions of preferentiality through time. This can easily be implemented. Furthermore, we may wish to set $w_{i,j,l,m} = 0$ for certain values of the subscripts to see if the effects of covariates and/or the effects of preferential sampling differs between the initial site placement process and the site retention process.

Clearly the above modeling framework has potential for over--fitting and model non--identifiability among others things. Thus careful choice of prior distributions, linear constraints on the latent effects (e.g. sum--to--zero constraints) and exploratory analysis is vital to fully utilize this model framework.

\section{Case study: the data}

\begin{figure}[H]
\begin{center}
\includegraphics[scale = 0.5]{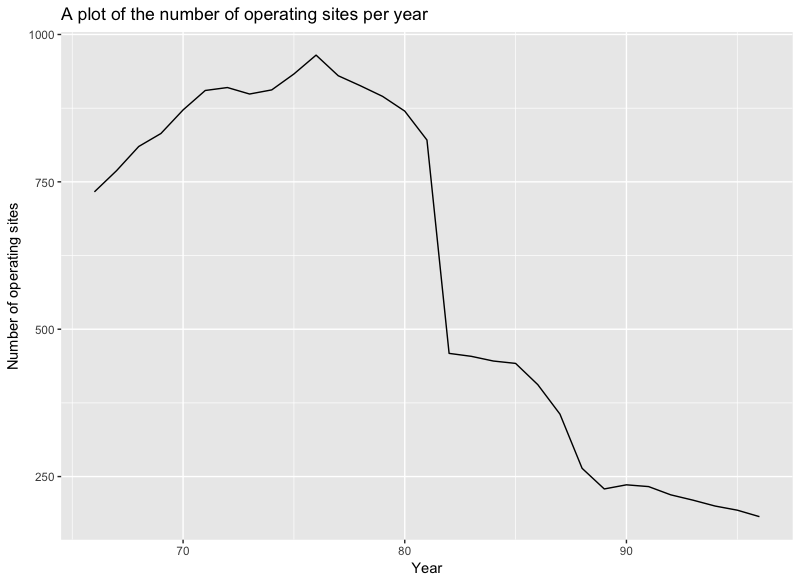}
\caption{A plot showing the number of the monitoring sites that are operational at each year and have data capture of at least 75\%. Note that a total of 1466 sites were operational at some point in time. }
\label{fig:Number_sites_operating_plot}
\end{center}
\end{figure}

Annual concentrations of BS were obtained from the UK National Air Quality Information Archive (www.airquality.co.uk). Set up in 1961, this was the world's first coordinated archive of national air pollution monitoring networks. While it was being established the network increased in size and the initial growth was quite rapid; from 800 sites in 1962, 1159 sites in 1966 to 1275 sites in 1971 (see Fig \ref{fig:Number_sites_operating_plot}). After this initial period the overall size of the network declined due to rationalisation and in response to changing levels of air pollution; in 1976 there were 1235 operational sites, 563 in 1986, 225 in 1996 and 65 in 2006. 

Site locations (at a 10 m resolution) and annual average concentrations of BS ($\mu g m^{-3}$) were obtained from monitoring sites. For the reasons given  by \citet{shaddick2014case}, we restrict ourselves to only the sites operating between April 1966 and March 1996 and with data capture of at least 75\%, equivalent to 273 days a year (as stated in the EC directive 80/779/EEC \cite{colls02airpollution}). The locations of all these sites (i.e the population ${\cal{ P}}_1$ considered in this paper) can be seen in Fig \ref{fig:Sites_plot_P1}.  It can be seen immediately that a high density of sites are located near many major industrial cities such as London and the Midlands, with almost no sites located in the relatively sparsely populated north of Scotland. 

\begin{figure}[H]
\begin{center}
\includegraphics[trim={0.7cm 0 0 0},clip,scale = 0.6]{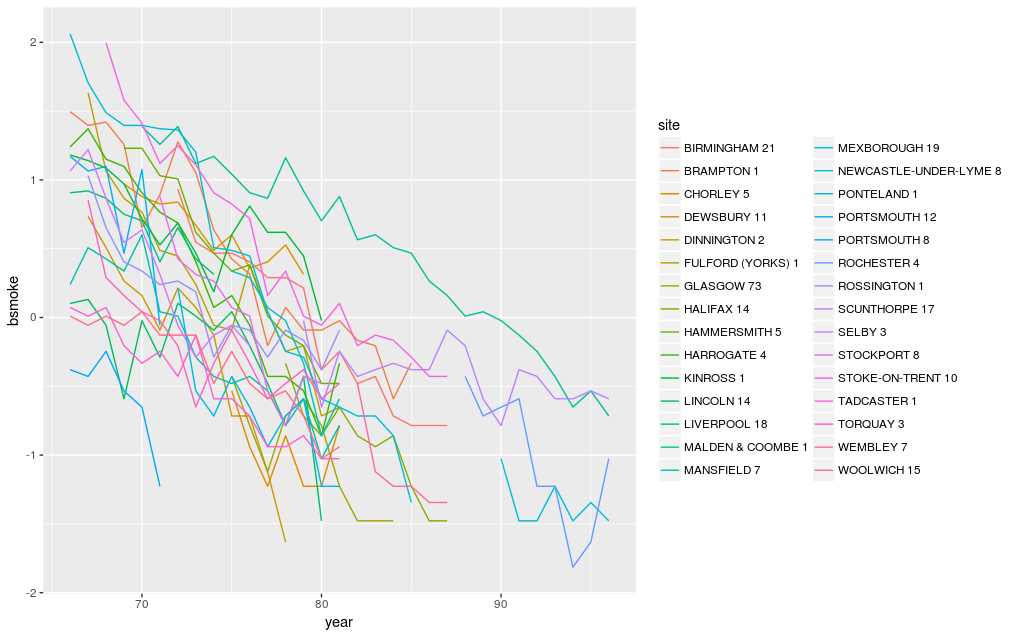}
\caption{A plot showing the mean black smoke level on the log transformed scale for 30 randomly chosen sites. Missing line segments indicate the site was offline that year.}
\label{fig:Spaghetti_plot}
\end{center}
\end{figure}

The decline in concentrations during this time period was most dramatic. Annual recorded network
means fell from 80 $\mu gm^{-3}$ in 1966 to 31 in 1976, 19 in 1986, 9 in 1996 and 5 $\mu gm^{-3}$ in 2006. Fig \ref{fig:Spaghetti_plot} shows a random sample of site--specific log--transformed annual BS levels. Concentrations of BS were typically highest in areas where the use of coal for domestic heating was relatively widespread, such as in parts of Yorkshire and within large cities.

Along with these large changes in concentrations, the dramatic changes in the size of the network can be seen in Fig \ref{fig:Number_sites_operating_plot} which shows the number of operational sites with at least 75\% data capture vs. year within the chosen study period. The initial increase in the size of the network can clearly be seen followed by the long--term reduction in the number of sites over time. Also evident is the marked reduction of the network in the early 1980s when there was a dramatic reduction in the number of sites of almost 50\% as the network was reorganised owing to falling urban concentrations. With such a dramatic drop in the size of the network, one must ask how the network reduction was chosen. Fig \ref{fig:Spaghetti_plot} shows a plot of a random sample of 30 sites' (log-transformed) black smoke trajectories. From this plot there appears to be evidence that the sites that remained in the network until the end were those providing the highest measurements. Thus we can see clear evidence for a response--biased network reduction process (i.e PS).

Thus we have a dataset that exhibits three interesting features: \begin{enumerate}
\item A high density of monitoring sites near major industrious regions, and hence near potential sources of BS. Conversely, an under-representation of the rural areas of Northern Scotland, Wales and Cornwall (Fig \ref{fig:Sites_plot_P1}), and hence areas with low expected BS.
\item A large change in concentrations of BS throughout the period of study, resulting in a rapidly evolving latent spatio--temporal process  (Fig \ref{fig:Spaghetti_plot}).
\item A network whose size dramatically changes through time (Fig \ref{fig:Number_sites_operating_plot}).
\item A network that underwent a biased redesign through time (Fig \ref{fig:Spaghetti_plot}), with the sites providing the smallest BS readings being dropped from the network.
\end{enumerate}
These four features provide the perfect opportunity for the model framework to both detect and attempt to correct for the effects of preferential sampling made within the network. In particular, depending on our choice of $\cal P$, we are investigating whether or not informative dropout/inclusion occurred in the operational network $S_t$ through time, and/or whether the network of observed sites is representative of Great Britain (GB) as a whole.

Note that the same exploratory analysis was conducted as in \citet{shaddick2014case}, and a quadratic temporal effect was found suitable to both fit the data and also provide a non--complex relationship to explain the observed decline in (log transformed) concentrations over time. Variograms were constructed for each year separately and for the average over all years, both on the original data and on the residuals from the temporal model; a spatial model from the Matern class seemed an appropriate choice. 

\section{Modelling}

We build one model from the general framework introduced in Section 2. We fit and present the results from three implementations of this model to display the features of the modelling framework. The three implementations are developed through a combination of imposing strict constraints on the PS parameters (i.e. by imposing point mass priors on the $\textbf{d}$ parameter vector), and changing the population under consideration. These three implementations clearly demonstrate the ability of the model framework to both detect, and adjust for, PS. Furthermore, they highlight the components of the model involved with the PS detection and correction, and help to demystify the method away from being a black--box approach. 

The joint model developed incorporates the effects of selection by sharing the random effects present in the observation process with the site--selection process. In particular, the selection process is allowed to use information from both spatially varying Gaussian processes and spatially--uncorrelated site--specific effects, to determine the site selection probabilities each year. If preferential sampling is detected, then this model should help to de--bias predictions of the ${\cal{ P}}_1$ and ${\cal{ P}}_2$--means relative to those reported from the raw data, by moving their point predictions against the direction of preferentiality. The magnitude of this movement is dependent upon: the flexibility of the model, the magnitude of the estimated PS parameters $d_\beta, d_b$, and the choice of $\cal P$. This fact is clearly demonstrated by the results from the three implementations.

The \textbf{same} joint model, and computational mesh is used across all three implementations. The differences seen in the results come only from the different assumptions placed upon the site--selection processes and populations. In the first implementation, the site--selection process is forced to be independent from the pollution process in the first implementation through the point mass prior at 0 imposed on $d_\beta, d_b$. In other words we constrain the PS parameters to be zero. Consequently the subsequent inference from this model will ultimately be equivalent to the inference from a model without any site--selection process component. In the second and third implementations, we remove this constraint, and two different choices of $\cal P$ are made to address two alternative scenarios. 

All modelling is performed in R-INLA with the SPDE approach \citep{lindgren11,rue2009approximate,rue2017bayesian}. This enables the rapid computation of approximate Bayesian posterior distributions for both the model variables and latent effect predictions. It does this by approximating the spatio--temporal processes with a Gaussian Markov random field (GMRF) representation by solving an SPDE on a triangulation grid. Details can be found in \citet{lindgren11}. Due to the large size of the dataset and the desired spatial prediction, MCMC approaches without sophisticated approximations would be infeasible. This is due to the computationally expensive operation of inverting large, dense spatial covariance matrices being required at each MCMC iteration to evaluate the likelihood. The SPDE approach, by developing a GMRF representation to the spatial fields, only requires the computationally cheaper operations of computing the inverse and the determinants of sparse precision matrices -- a task that is made possible with numerical sparse matrix libraries.   

\subsection{Data cleaning}
A few data cleaning steps were carried out before fitting the modelling. Due to the right skewness of the black smoke observation distribution, we applied the natural logarithmic transformation to the values to make the observation distribution more Gaussian in shape. Since the natural logarithm is a non--transcendental function, meaning in particular that its series representation contains an infinite series of powers of its argument, we first divided each value by the mean of all the recorded black smoke levels to make the response dimensionless. This ensures not only that the inference remains valid, but also readily interpretable as they are in effect compared to a natural origin. Next, we scaled the Eastings and Northings coordinates by the standard deviation of the Eastings, and re--scaled the years to lie in the interval [0,1] to stabilise the temporal polynomials used in later analysis.

\begin{figure}
\begin{center}
\includegraphics[scale=0.9]{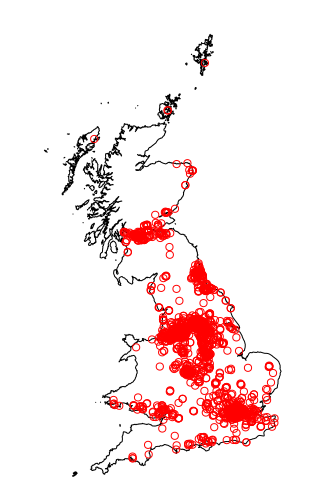}
\caption{A plot of Great Britain, with the locations of the observed sites, and hence ${\cal P}_1$ shown.}
\label{fig:Sites_plot_P1}
\end{center}
\end{figure}

\subsection{Observation Process}
The following model for the observation process is used for all three implementations seen shortly. The specification follows from \citet{shaddick2014case} and is formulated as follows. Let $Y_{i,j}$ denote the observed log black smoke ratio at site $i$, situated at $\textbf{s}_i$, at time $t_j$ $i \in \{1,...,M\}, j \in \{1,...,N\}$. Let $t^{\star}_j$ denote the $j^{\rm th}$ time--scaled observations that lie in the interval $[0,1]$. Let $R_{i,j}$ denote the random selection indicator for site $i$ at time $t_j$. Let $R_{i,j} = 1$ or $0$ depending on whether or not the site was operational in that year and provided the minimum number of readings outlined earlier. Note that there are 1466 sites that record at least one annual reading, and $N = 31$. 

\begin{align}
\left( Y_{i,j} | R_{i,j} = 1 \right) &\backsim \textnormal{N}\left(\mu_{i,j}, \sigma^2_\epsilon\right) \nonumber \\
\mu_{i,j} &= (\gamma_{0} + b_{0,i} + \beta_{0}(\textbf{s}_i) ) + ( \gamma_{1} + b_{1,i} + \beta_{1}(\textbf{s}_i))t_j^{\star} + (\gamma_{2} + \beta_{2}(\textbf{s}_i))(t_j^{\star})^2 \nonumber \\
\left[\beta_{k}(\textbf{s}_1), \beta_{k}(\textbf{s}_2), ..., \beta_{k}(\textbf{s}_m) \right]^T &\backsim^{IID} \textnormal{N}\left(\textbf{0}, \Sigma(\zeta_k)\right) \hspace{0.2cm} \textnormal{for} \hspace{0.2cm} k \in \{0,1,2\} \nonumber \\
\left[b_{0,i}, b_{1,i}\right] &\backsim^{IID} \textnormal{N}\left(\textbf{0}, \Sigma_b \right) \hspace{0.2cm} \Sigma_b = \begin{bmatrix}
\sigma^2_{b,1} & \rho_b \\ \rho_b & \sigma^2_{b,2}
\end{bmatrix} \nonumber \\
\Sigma(\zeta_k) &= \textnormal{Matern}\left(\zeta_k\right) \nonumber \\ 
\theta = \left(\sigma^2_\epsilon, \gamma, \zeta_k, \sigma^2_{b,1}, \rho_b\right) &\backsim \textnormal{Priors}. \nonumber
\end{align}

The choice of the observation process model is explained as follows. The sources of variation can be broken up into three components: global variation, independent site--specific variation and smooth spatially correlated variation. To ensure model identifiability, we enforced sum--to--zero constraints on all random effects ($\beta$ and $b$), and furthermore we did not estimate spatially--uncorrelated random effects $b$ at locations with no observations. Note that in the notation of Section 2, the $b$ and $\beta_k(\textbf{s}_i)$ terms are a examples of the $\beta(\textbf{s}, t)$ latent effects and thus $q_1 = 5$. For readability we choose to separate the notation for these effects. Note that, whilst the $b$ terms are assumed independent between sites, the terms $b_{0,i}, b_{1,i}$ are assumed a-priori to be a realisation from a (possibly-correlated) multivariate Gaussian distribution with covariance matrix $\Sigma_b$. 

The global temporal trend is captured by the $\gamma_k$ terms since these parameters remain constant across the sites. As in \citet{shaddick2014case},  when comparing various models for the first (non-joint) implementation, more complex temporal relationships (such as splines) were not favoured by multiple model selection criteria including DIC. Secondly, the independent site--specific variations are captured by the IID random intercepts and random slopes ($b_{0,i}, b_{1,i}$). In geostatistical terms, the $b$ terms act as nugget effects for their corresponding $\beta_k(\textbf{s})$ terms. The (nugget--free) $\beta_k(\textbf{s})$ terms then capture the smooth spatially--correlated variation. Models without the $b$ terms showed large residual site--specific errors. Thus it appears that small--scale factors may be a large source of variability in the measured black smoke trajectories, independent from the regional location alone. Note that separate spatially--correlated Gaussian fields for each year were tested (i.e using a separate $\beta_{0,j}(\textbf{s})$ field for each year), but did not improve the model fit. 

The intuition behind the short scale $b$ terms in the model is as follows. An observation tower close to a large source of black smoke (e.g. a road, a polluting factory or a power station) would likely yield a much higher annual reading than placing it say half a kilometer away from such a source. Since this spatial scale is much smaller than that captured by the $\beta_k(\textbf{s})$ processes, these differences will not be accounted for without either including covariates that capture the causes of these effects (e.g. distance from the nearest pollutant source), or by allowing each site to have it's own deviation from the smoothly predicted field via either a fixed or random, site--specific effect. Note that spatially--uncorrelated random quadratic slopes $b_{2,i}$ were not found to improve the model fit  with respect to DIC under the first implementation and actually led to a large instability in the predictions of sites that took fewer measurements. It appears that the inclusion of these terms led to some over--fitting.  

The choice of priors for the hyperparameters $\theta$ were made to make them as weakly informative as possible and hence to reduce their effects upon the posterior results, but also to bound their values inside sensible limits. Despite the fact that previous analyses have been made on this dataset, we only use vague information from these results when constructing the priors. We discuss the details of the chosen priors in the supplementary material.

\subsection{Site--selection Process}
The following model for the site--selection process is used for all three implementations with the aim of emulating the complex decision--making processes that occurred when setting up the monitoring network.  Let: $R_{i,j}$ denote the random selection indicator for site $i$ at time $t_j$; Let $R_{i,j} = 1$ or $0$ depending on whether or not the site was operational in that year and provided that the minimum number of readings outlined earlier is attained.  Let $r_{i,j} \in \{0,1\}$ denote the realisation of $R_{i,j}$ for site $i$ at time $t^{\star}_j$, $i \in \{1,...,M\}, j \in \{1,...,N\}$. Finally, $\textbf{s}_i$ denotes the location (the scaled Eastings and Northings coordinates) of site $i$. The model is then: 

\begin{align}
R_{i,j} &\backsim \textnormal{Bernoulli} \left(p_{i,j}\right) \nonumber \\
\textnormal{logit} p_{i,1} &= \alpha_{0,0} + \alpha_1 t^{\star}_1 + \alpha_2 (t^{\star}_1)^2 + \beta^{\star}_1(t_1) \nonumber \\ 
&+ \alpha_{rep} I_{i,2} + \beta^{\star}_{0}(\textbf{s}_i) \nonumber \\
&+ d_b \left[b_{0,i} + b_{1,i}(t^{\star}_{1})\right] + d_\beta \left[\beta_{0}(\textbf{s}_i) + \beta_{1}(\textbf{s}_i)(t^{\star}_{1}) + \beta_{2}(\textbf{s}_i)(t^{\star}_{1})^2 \right]  \nonumber \\
\textnormal{for} \hspace{0.2cm} j \neq 1 \hspace{0.4cm} \textnormal{logit} p_{i,j} &= \alpha_{0,1} + \alpha_1 t^{\star}_j + \alpha_2 (t^{\star}_j)^2 + \beta^{\star}_1(t_j) \nonumber \\ 
&+ \alpha_{ret} r_{i,(j-1)} + \alpha_{rep} I_{i,j} + \beta^{\star}_{0}(\textbf{s}_i) \nonumber \\
&+ d_b \left[b_{0,i} + b_{1,i}(t^{\star}_{j-1})\right] + d_\beta \left[\beta_{0}(\textbf{s}_i) + \beta_{1}(\textbf{s}_i)(t^{\star}_{j-1}) + \beta_{2}(\textbf{s}_i)(t^{\star}_{j-1})^2 \right]  \nonumber \\
I_{i,j} &= \mathds{I} \left[ \left(\sum_{l\neq i} r_{l,j-1} \mathds{I}\left(||s_{i} - s_{l}|| < c\right) \right)  > 0 \right] \nonumber \\
\left[\beta^{\star}_{0}(\textbf{s}_1), ..., \beta^{\star}_{0}(\textbf{s}_m) \right]^T &\backsim \textnormal{N}\left(\textbf{0}, \Sigma(\zeta_R)\right] \nonumber \\
\Sigma(\zeta_R) &= \textnormal{Matern}\left(\zeta_R\right) \nonumber \\ 
\left[ \beta^{\star}_1(t_1),..., \beta^{\star}_1(t_T)  \right]^T &\backsim \textnormal{AR1}\left(\rho_a, \sigma^2_a\right) \nonumber \\ 
\theta_R = \left[\alpha, d_b, d_\beta, \rho_a, \sigma^2_a, \zeta_R \right] &\backsim \textnormal{Priors}. \nonumber
\end{align}
\\
 
The first rows of the linear predictors comprise the global effects of time on the log odds (and thus eventually the probability) of selection. We allow for a quadratically changing global log odds of selection with time, and allow for a global first--order autoregressive deviation from this quadratic change (denoted by $\beta^{\star}_1(t_j)$). This term represents the change in time of both the political and public moods regarding the need for maintaining the overall network size. New governments may well prioritise public spending on the environment in different ways and furthermore, the public's approval of environmental spending likely changes in light of new knowledge. Additionally, large changes in the size of the public monitoring network can be seen around 1982 (see Fig 2). Here a sharp decrease in the size of the network occurred, reducing the number of sites by almost half. The smooth quadratic effect of time clearly would not suffice to capture this short term trend and thus a random effect seems compelling, especially one that is able to adequately capture this short term change (i.e. overdispersion), such as the autoregressive term we used.

The second rows of the linear predictors represent the site--specific factors influencing the log odds ratio in favour of a site's inclusion in the network $S_j$ at time $t_j$. Firstly, $\alpha_{ret}$ represents what we call the ``retention effect''. This term reflects how the probability a site is selected in a given year, changes conditional upon its inclusion in the network in the previous year. Since large costs can be incurred in setting up monitoring sites at new locations, it is plausible that network designers would favour the maintenance of existing sites over their replacement at new site locations, even if the conditions at other sites (represented by the other terms in the linear predictor) are more favourable. In fact, it is this indicator variable that determines whether or not the linear predictor corresponds to the site-placement process or the site-retention process. If we wanted to investigate the possibility that the effects of PS or covariates were different between the two processes, then we could include additional product terms between the various effects and $r_{i,j-1}$ to capture this change. Here, we share all parameters across the two processes and allow only a unique intercept to exist between the processes. This is discussed in depth later.  

In contrast, $\alpha_{rep}$ captures the repulsion effect. $I_{i,j}$ denotes an indicator variable that determines whether or not another site in the network placed within a distance $c$ from site $i$ was operational at the previous time $t_{j-1}$. Plausibly network designers would not want to place sites close to an existing site. 
Conversely, there may be unmeasured regional confounders affecting the localised site--selection probabilities (e.g. population density) that may lead to additional clustering that cannot be explained by the model without the inclusion of the confounder. This parameter should help to capture any additional clustering that may be present. We choose the hyperparameter $c$ to be 10km.

Finally, there may be a larger motivation to place more/fewer sites in certain areas of the UK throughout $T$, that cannot be explained by the other terms in the model. This may be due to population density or due to increased/decreased political incentives in this area. We attempt to capture such spatially--varying area effects in the $\beta^{\star}_0(\textbf{s})$ field. This can be viewed as a spatially--correlated correction field similar to that used by \cite{pati2011bayesian}. Note that this is fixed in time with the aim of avoiding identifiability issues.

Whilst it may appear that we have included a lot of effects in the site--selection process, it is of paramount importance to adequately capture and remove as many sources of variability from the site--selection process as possible. The preferentiality parameters should therefore only act upon the \textbf{residual} signal, after such effects have been removed. Since we are dealing with a large quantity of spatio-temporal data, we are able to learn the temporal features affecting site-selection and thus we can attempt to emulate the true process itself. This is in stark contrast with the spatial setting. By removing large sources of variability from the site--selection process first, we reduce the risk of over--estimating the stochastic dependence between the selection and observation processes and hence reduce the risk of over-adjusting our parameter estimates and predictions.

The third and final rows of the linear predictor represent the preferentiality parameters of the selection process, following the work of \cite{diggle2010geostatistical}. We decide to separate the preferentiality into two sources: small--scale deviations from the localised average black smoke levels, and the medium--scale regional deviations from the UK--wide annual black smoke levels. In recognition of the landmark paper by Diggle, we denote the two parameters by $d_b, d_\beta$ respectively. Since we have constrained both the $[b_{0,i}, b_{1,i}]$ terms and the $\beta_k(\textbf{s})$ processes to sum to zero, the terms being multiplied by $d_b, d_\beta$ represent deviations from the $\cal P$--mean. Both of these effects are allowed to affect site selection independently. The interpretation of these PS parameters depends largely upon the choice of the population $\cal P$. All PS effects detected are \textbf{after} controlling for the other site--selection effects.

In consideration of the discussions following \cite{diggle2010geostatistical}, for $j > 1$ site selections made at time $t_{j}$ involve estimated black smoke levels based on observations made at the previous time $t_{j-1}$. Thus in our model we do not assume the network designers formulate sight selection decisions based on black smoke forecasts into the future or for the current unobserved year, but on predicted quantities at the previous time step. Therefore in our framework, we model the site--selection as being reactive for times $t_j : j > 1$. Using the notation from 2.2, $\phi_{i,l,k}(t_j) = t_{j - 1} \forall i,l,k$ and $ t_j > 1$. If the true selection mechanism is believed to be different, then the change of paradigm is trivial. For computational savings, we base the site selection at time 1 to be based on the estimated field at time 1 (i.e. $\phi_{i,l,k}(t_1) = t_1$). Our choice of priors are discussed in depth in the supplementary material.

\subsection{Three implementations}

For Implementation 1 we constrain the PS parameters $d_b, d_\beta$ to equal 0. Thus Implementation 1 incorporates the prior assumption that no stochastic dependence between the site--selection process and the observation process was present and thus that no PS occurred. A direct result of this independence assumption is that the posterior distribution of the observations process $Y$ is the same, regardless of the specification of either the site--selection model terms, or the choice of the population of sites $\cal P$ to consider for selection. Thus the results from Implementation 1 will simply match the typical spatio--temporal analyses conducted in practice, ignoring site--selection. This will be used as our baseline for comparison. 

For Implementation 2, we remove the zero constraints on the PS parameters, imposing instead weakly informative Gaussian priors with mean 0 and variance 10. For Implementation 2, we consider \textbf{only} the 1466 observed site-locations for selection at each time $t \in T$. We define this as Population 1, ${\cal{P}}_1$ and thus $M = |{\cal{ P}}_1| = $ 1466. Population 1 is shown as the red circles in Fig \ref{fig:Sites_plot_P1}.  

\begin{figure}
\begin{center}
\includegraphics[scale=0.25]{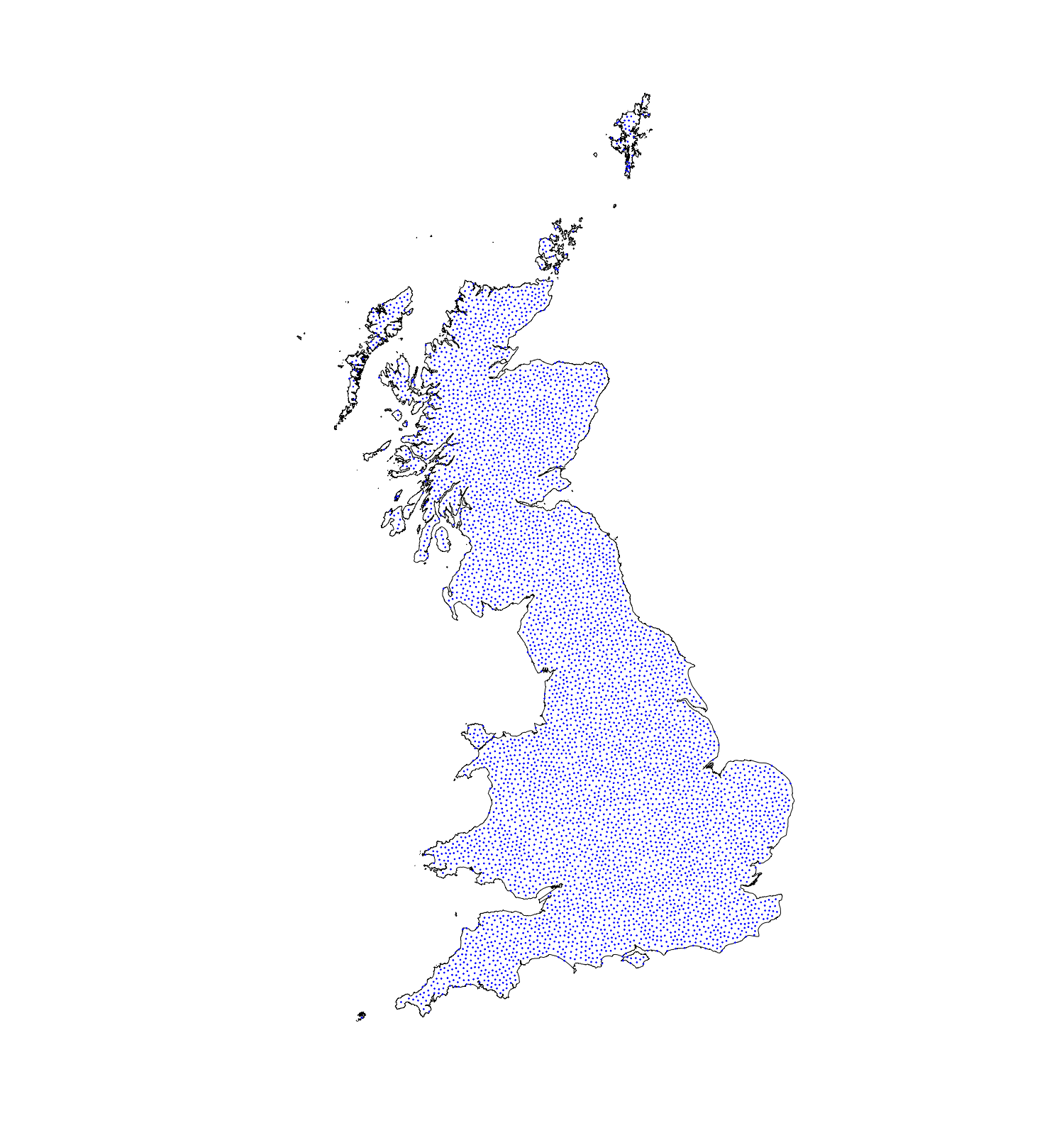}
\caption{A plot of the locations of \textbf{all} sites considered for selection in Population 2. The locations are shown as blue dots, many of which are in regions of low human population density.}
\label{fig:Sites_plot_P2}
\end{center}
\end{figure}

For Implementation 3, we replace the zero constraints with the same Gaussian priors, but consider a different population of sites for selection at each year, ${\cal{P}}_2$. For ${\cal{P}}_2$ thousands of pseudo--sites are also considered for selection at each time step along with the observed sites from ${\cal{P}}_1$. We ensure the locations of the pseudo--sites are uniformly distributed throughout Great Britain (GB) and placed with high density. It has been shown that estimates and corresponding standard errors of all (non-intercept) parameters converge toward those of the equivalent inhomogeneous Poisson point process as the number of pseudo--sites tends towards infinity, so long as the density of the points is uniform (in probability) \citep{warton2010poisson,fithian2013finite}. Thus there is some duality with the approach of DMS \citep{diggle2010geostatistical} and our Implementation 3. The locations of ${\cal{P}}_2$ are shown in Fig \ref{fig:Sites_plot_P2}.

For Implementation 2 we aim to see if the network \textbf{evolved} preferentially. That is, out of the observed sites, were sites added and dropped from the network in a manner that was dependent upon the value of the latent black smoke process and hence missing not at random (MNAR). Under Population 1, since we do not consider locations within the unsampled regions for selection, no additional information is being added to the unsampled regions. Hence we do not expect the estimates of BS to change much at these locations unless estimates of the site--trajectories and hence the ${\cal{P}}_1$--mean change. Furthermore, we are unsure if the joint model will substantially adjust estimates of the ${\cal{ P}}_1$--mean, even if PS is detected. This is since results from a small simulation study we conducted suggest that if we have a case where we fit an inflexible temporal model to a dataset whose sites have a long average consecutive lifetime, estimates will remain largely the same due to the over--determined nature of the problem. In fact, the sites in the dataset provide an average of 12 consecutive years of readings, with the minimum consecutive lifetime of a site being 6 years. Additionally the deviation from the quadratic trend is typically small (Fig 3). Thus we may expect only a small change to the results seen from Implementation 1.  

For Implementation 3 we investigate if the network of operational sites at each time $S_t : t\in T$ is being located throughout GB ($\cal S$) in a preferential manner. Thus the interpretation of preferential (i.e. response--biased) network evolution is lost under this choice of population. Instead, these PS parameters $d_\beta , d_b$ now measure the degree to which the operational network ($S_t$) is preferentially located in $\cal S$ through time $\cal{T}$. This is due to our second population ${\cal{ P}}_2$ covering $\cal S$ uniformly and hence considering each point $\textbf{s} \in \cal S$ as being equally likely to be sampled a-priori. This is unlike Population 1, which didn't include large areas of unsampled Scotland, Wales and Cornwall for selection at each time $t \in T$. Thus Population 2, by adding additional information to the unsampled regions via the site-selection process, should inform the joint model about the appropriate adjustment of BS estimates in the unsampled regions according to the nature of PS detected. Put differently, the joint model will extrapolate any associations detected between the site--selection process and the underlying latent effects into the unsampled regions. 

In fact, hidden away in the details of Implementation 3 is the fact that the Bernoulli random variable models two processes simultaneously. Implementation 3 can be considered as being a joint model with three processes: an observation process, an initial site--placement process and a site--retention process. The latter two are fit using only one Bernoulli likelihood. The initial site--placement process is fit using a conditional logistic regression approximation to a log-Gaussian Cox process, and is similar to that seen in \cite{diggle2010geostatistical}. The site--retention process is modeled as a Bernoulli random variable. Inside the linear predictor of the Bernoulli likelihood, the indicator variable $r_{i,(j-1)}$ points the linear predictor towards the site--placement process when it is equal to 0 or towards the site--retention process when it is equal to 1. In our example, we only allow for a unique intercept to exist across the two processes, sharing the remaining parameters. Thus we assume that the effects of all the covariates and the effects of PS are constant across the two processes. This assumption can be relaxed by including interaction effects between $r_{i,(j-1)}$ and the other parameters, including the PS parameters. 

Note that care is required to ensure that only the pseudo--sites contribute a zero to the Bernoulli likelihood for the site--placement process across all years. Furthermore, for our application, we must ensure that only the sites that have been removed from the network in year $j$ contribute a zero to the Bernoulli likelihood for the site--retention process at year $j$. This ensures that no site in the network was ever re-installed after its removal, a fact seen in our data. Clearly then, the choice of zeros here is application--dependent. Additional details are given in the supplementary material.   

The ability of our joint model to adjust estimates of the pollution process at a point $\textbf{s}$ depends upon the distance of the point from the nearest monitoring site in the network. For pseudo--sites further from an observed site than the effective range of the spatially varying $\beta$ processes, essentially all the degrees--of--freedom of the spatially--varying quadratic terms $\beta_k(\textbf{s})$ are available for use in fitting the site--selection process to make the posterior probability of repeated non-selections (i.e. the $r_{i,j} = 0$'s) of the pseudo--site high. Since we have no black smoke observations here, the fitting of the quadratic slopes to these pseudo--sites is therefore an under--determined problem. Thus we would expect the estimates of black smoke here to be different. For pseudo--sites very close to an observed site (i.e. well within the effective range), we would expect the estimates at the pseudo--site locations to remain largely unchanged, since the problem remains over--determined. For pseudo--sites within the effective range of, but not immediately next to an observed site, we expect estimates to change moderately since the problem is weakly-determined.

\subsection{Model identifiability issues}
When fitting a model this large, issues around model identifiability commonly arise, namely the possibility of the data providing information about the model parameter values through the likelihood. We assessed these issues with two approaches. First, we enforced sum--to--zero constraints on all the random effects to ensure they are simply localised deviations about a global trend. As discussed in the supplementary material, we placed PC priors \citep{simpson2017penalising,fuglstad2017constructing} on the Matern parameters of the Gaussian processes to provide some prior information on the range and scale, while reducing the possibility of overfitting the data. 

To confirm that we had fully resolved the model identifiability issues, we then conducted a small simulation study. We sampled the data from various models similar in form to the joint model introduced in sections 4.2 and 4.3 to see if the posterior estimates of both the parameters and the space--time field covered the true values. Interestingly, for a much smaller dataset, we found no identifiability issues except for the range parameter on the $\beta_{0}^{\star}(\textbf{s})$ process. Here the mean squared error of the point estimates of this parameter were very high relative to the other parameters, although the nominal coverage levels and bias remained good. This could be a sign of identifiability issues surrounding this effect, or perhaps could be due to the difficulty with estimating a Matern field using only small amounts of binary point data. All other parameter estimates in the simulation studies, as well as posterior predictions were good. Of most interest was the model's capability to detect the preferentiality parameters $d_\beta, d_b$ with high precision, negligible bias and with posterior credibile intervals attaining nominal coverage levels. 

Interestingly, we experience the same difficulties with identifying the $\beta_{0}^{\star}(\textbf{s})$ process in our case study. Our estimated marginal distributions for the range parameter of the $\beta_{0}^{\star}(\textbf{s})$ process in the UK black smoke case study were all found to have 95\% posterior credible intervals all around (0.03, 1.18). Given that we scaled the coordinates, this range of estimates covers a range of distances from very small up to very large. Hence it appears the model encounters difficulties with estimating this parameter. Importantly, the posterior means of the standard deviation of this effect were around 0.03 with 95\% credible intervals lying in the region of between 0.00 and 0.08. Thus ultimately this effect has minimal impact upon the model fit.  

We also assessed the ability of the joint model framework under simulated PS settings to de-bias estimates of site--specific trajectories and network averages (equivalent to the ${\cal{ P}}_1$--mean). Two such simulation studies considered distinct temporal trends. The first fixed the temporal component to be rigid, the second allowed for a flexible nonlinear trend. In particular, we witnessed that under a rigid (spatially-varying) linear slopes model, when the average of the consecutive lifetimes of the sites is high, the bias induced in the site--specific estimates and the ${\cal{ P}}_1$--mean that occurs from ignoring the site-selection process is almost zero. This is due to the problem of being over-determined -- only a few observations of the process at each site are required for the model to accurately forecast/backcast estimates throughout $T$. This is similar to what is seen in the case of the UK black smoke dataset. Conversely, when the temporal trend is highly nonlinear and the average consecutive lifetimes of the sites are short, the biases in parameter estimates, site--specific predictions and estimates of the ${\cal{ P}}_1$--mean through time can all be high if we ignore the site-selection process. This phenomena is well understood in the joint longitudinal mixed models literature -- the higher the measurement error and the more nonlinear the subject--specific trajectories, the more inference can change under a joint model. To provide a `highly nonlinear' trend, we opted to use an independent realisation of a Matern field for each of the 30 simulated `years'. The insights from these two scenarios help explain the results seen shortly in Implementation 2. They also hint that changes to inference under Implementation 2 ${\cal{ P}}_1$ would be highest for applications with mobile monitoring sites.

\section{Results}
We focus our attention upon the following issues and objectives: \begin{enumerate}
\item Do implementations 2 and/or 3 detect that, within the network of observed sites (i.e. Population 1), the sites have been preferentially added and removed even after controlling for the various covariates included in the site--selection process? If so, has this been done based upon short--range, site--specific deviations from the regional mean black smoke, and/or medium-range regional deviations from the annual $\cal P$-- mean?
\item When considering Implementation 3, does the model detect that the network of operational sites $S_t$ have been preferentially located within GB ($\cal S$) through time, even after controlling for the various covariates included in the site--selection process?
\item Do estimates of the black smoke annual means in GB (i.e. the ${\cal{P}}_2$--mean) change significantly when we consider the stochastic dependence between the placement of the sites and the black smoke field? 
\item If we backcast and/or forecast the predictions at all observed site locations (i.e. $\textbf{s} \in {\cal{P}}_1$) at all times, how do the estimated black smoke levels differ between the operational ($S_t$) and offline sites ($S_t^C$)? Do these differences change in time, and if so, does the apparent priority of site placement change through time? 
\item Given the original purpose of the air quality network for monitoring the progress achieved by the Clean Air Act in reducing the population exposure levels to both black smoke and sulphur dioxide \citep{McMillanUnpublishedAirQualityHealthEffects}, if we average the estimated black smoke field across Great Britain's population, do the estimated population-average exposure levels change between the implementations?
\item Considering the 1980 EU black smoke guide value of 34 $\mu$gm$^{-3}$, how does the estimated proportion of GB exceeding this value change through time? What are the differences across the three implementations? Furthermore, how do estimates of the proportion of the population exposed to BS levels above this value change under the three implementations? 
\end{enumerate}

In this section we refer to some secondary plots found in the supplementary material. When this occurs we will put a $\star$ superscript above the figure number (e.g. Fig \ref{fig:naive_map_plots}$^\star$). 

\begin{table}[h!]
\centering
 \begin{tabular}{||c c c c||} 
 \hline
 Parameter & Implementation 1 & Implementation 2 & Implementation 3 \\ [0.5ex] 
 \hline\hline
 $d_\gamma$ & 0 (0) & 0.62 (0.17) & 2.77 (0.01)  \\ 
 $d_b$ & 0 (0) & 0.06 (0.04) & 0.12 (0.01) \\
 \hline
 $\beta_0$ & 96.50 & 94.94 & 21.87 \\
 (trans scale) & 1.15 (0.02) & 1.13 (0.01) & -0.34 (0.09) \\
 \hline
 $\rho_b$ & -0.77 (0.02) & -0.76 (0.02) & -0.78 (0.00) \\
 \hline
 $\alpha_{ret}$ & - & 6.18 (0.06) & 6.47 (0.06) \\ 
 $\alpha_{rep}$ & - & 0.08 (0.11) & 0.82 (0.10) \\ 
 [1ex] 
 \hline
\end{tabular}
\caption{A table showing the posterior mean and standard deviations for parameter estimates for the three implementations. Note that the top row estimates of $\beta_0$ have been transformed back onto the original data scale.}
\end{table}

\subsection{Implementation 1 -- assuming independence between $Y$ and $R$}

\begin{figure}
\begin{center}

\includegraphics[trim={0 0 2.7cm 0.7cm},clip,scale=0.5]{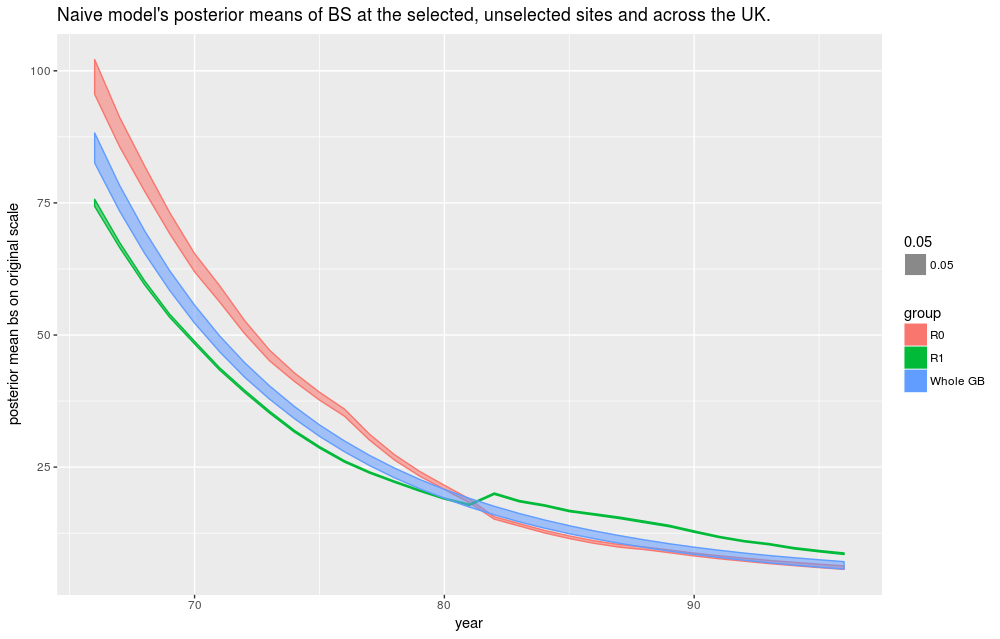}
\caption{Implementation 1. In green are the BS levels averaged over sites that were selected in ${\cal{P}}_1$ (i.e. operational) at time t. In contrast, those in red are the BS levels averaged over sites that were not selected in ${\cal{P}}_1$ (i.e. offline) at time t. Finally, in blue are the BS levels averaged across Great Britain. Also included with the posterior mean values are their 95\% posterior credible intervals. If printed in black-and-white, the green band is initially the lower line, the red band is the upper line and the blue band is initially the middle line.}
\label{fig:Post_Mean_Sites_Plot_Imp1}
\end{center}
\end{figure}

If we assume independence between $Y$ and $R$, the posterior results about the observation process $Y$ from Implementation 1 are identical to those that would have been discovered from fitting only the observation process (i.e. fitting only the $Y$ model). As expected, especially high values of black smoke are predicted to exist around the North West and Yorkshire areas of England in 1966. This area covers the major cities of Liverpool, Manchester, Leeds and Sheffield, all industry--heavy cities at the time under study. By 1996 the relative levels of black smoke in these areas are far reduced and exceeded by the Greater London area. Counter--intuitively however, the estimated black smoke levels in the Scottish Highlands, an area with almost no manufacturing or industry are predicted to be relatively high (see Fig \ref{fig:naive_map_plots}$^\star$) across all time periods. This is a direct consequence of the absence of monitoring sites in this area (see Fig \ref{fig:Sites_plot_P1}), along with a lack of informative covariates included in the observation process $Y$ for this region. 

A typical location in the unsampled regions of the Scottish Highlands, Cornwall and The Borders sees their distance to the nearest site in ${\cal{P}}_1$ typically exceeding the estimated spatial ranges of the random fields. Consequently, model--estimates in such areas essentially equal the average of the observed pollution levels (i.e. the ${\cal{P}}_1$--mean). This feature can immediately be seen to be problematic since it is likely that the true black smoke levels will be below the ${\cal{P}}_1$--mean in these regions. Similar effects are seen in Cornwall and the Borders. As well, large standard errors (i.e. posterior pointwise standard deviations) for the predicted black smoke levels are found in these regions due to their lack of monitoring sites (see Fig \ref{fig:naive_map_plots}$^\star$). 

Next, we consider the model--estimated black smoke levels for all the observed site locations (i.e. Population 1) in Fig \ref{fig:Post_Mean_Sites_Plot_Imp1} at every time point. To investigate Objective 4, for each $t \in T$ we split the observed sites into the operational sites $S_t$ and offline sites $S^{C}_t$. The set of operational sites $S_t$ are defined to be the sites in Population 1 that recorded the minimum number of observations that year. The set of offline sites $S^{C}_t$ are defined to be the sites in Population 1 that failed to record this minimum number of observations that year. Note that $S_t \bigcup S^{C}_t = {\cal{P}}_1$ and $S_t \bigcap S^{C}_t = \emptyset$.

Here we can see that from Implementation 1, that it appears the sites were initially placed in regions with below--average black smoke levels between 1966 -- 1980 (see Fig \ref{fig:Post_Mean_Sites_Plot_Imp1}). This is inferred from the posterior mean black smoke levels -- they are significantly lower for the operational sites compared with the estimated GB--average. The lack of additional information for the unsampled regions of GB makes the estimates in these areas equal to the ${\cal{ P}}_1$--mean and thus the GB--average is nearly identical to the ${\cal{ P}}_1$--mean. Over time, the posterior means for the black smoke levels at the operational and offline sites converge, before the direction of preferentiality changes in 1982. The latter was the year a major network redesign was initiated, removing almost half of the operational sites (see Fig \ref{fig:Number_sites_operating_plot}). Here we see strong evidence the sites that remained in the network after this redesign were in locations with black smoke levels above the ${\cal{ P}}_1$--mean. This is due to the posterior mean black smoke levels being significantly higher for the operational sites compared with the offline sites. 

Thus from looking at the results from Implementation 1 alone, we gain some insight about Issues 1 and 4. It appears that the sites were preferentially sampled in almost all time periods. Initially the operational sites appear to have been placed in regions with black smoke levels below the ${\cal{ P}}_1$--mean, before being placed in regions with levels above the ${\cal{ P}}_1$--mean after the major network redesign in 1982. These results are significant with respect to 95\% credible intervals. However, doubts have been cast about the predicted black smoke levels in regions of GB known to have little industry or population density -- two major sources of black smoke. Since these regions cover large percentages of the surface area of GB, the effect of over--estimating the predictions in these areas would be a marked increase in the estimated GB--average black smoke level. Implementation 3 attempts to rectify this problem by extending the definition of $\cal P$ into these regions.

\subsection{Implementation 2 -- ${\cal{ P}}_1$}

\begin{figure}
\begin{center}
\includegraphics[trim={0 0 0 1cm},clip,scale=0.4]{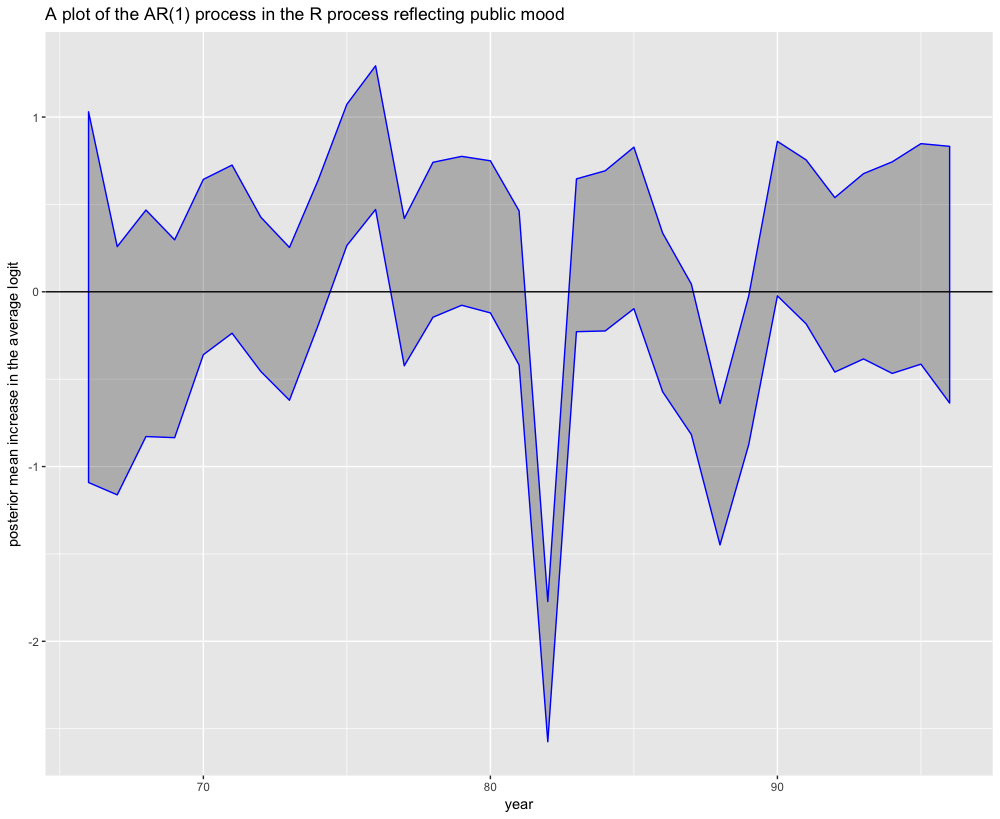}
\caption{A plot of the year--by--year change in the logit of selection captured by the autoregressive $\beta_1^\star(t)$ process in the $R$ process in Implementation 2. Note that the plot for Implementation 3 is almost identical.}
\label{fig:AR1_plot_imp2}
\end{center}
\end{figure}

Firstly, we consider the posterior parameter estimates for the two sources of preferentiality (see Table 1). These are denoted by $d_\beta, d_b$, the medium--range and short--range preferentialities respectively. Only the former effect $d_\beta$ was detected to be significantly nonzero with a posterior estimated value of 0.66 and a 95\% posterior credible interval of (0.34, 0.99). The posterior estimate of the short--range preferentiality was 0.06 with a 95\% posterior credible interval of (-0.01, 0.15). Thus in both cases the direction of preferentiality was positive, suggesting that year--by--year, the site placements are positively associated with the relative levels of black smoke at the site location, especially with the regional--average level. 

Interestingly however, despite this reasonably strong evidence of preferential sampling, the posterior predictions of black smoke levels are almost identical to those from Implementation 1. Fig \ref{fig:post_mean_sites_plots_imp2}$^\star$ and Fig \ref{fig:joint1_map_plots}$^\star$ both appear strikingly similar to those from Implementation 1 (Fig \ref{fig:Post_Mean_Sites_Plot_Imp1} and Fig \ref{fig:post_mean_sites_plots_imp2}). In particular, no obvious changes in the estimated BS levels are seen across the unsampled regions of the Scottish Highlands or the foot of Cornwall. Furthermore, the posterior mean black smoke level averaged across GB remains largely the same throughout time relative to the predictions from Implementation 1.

Thus it appears that despite the joint model detecting preferential sampling under Population 1, little--to--no change in the posterior estimates is seen in either the GB--average levels or the individual site--specific BS trajectories. This is in stark contrast with the observed de--biasing of the regional mean witnessed shortly under Implementation 3. The explanation for these two results may be best explained in terms of the two different populations ${\cal{ P}}_1, {\cal{ P}}_2$ of sites under consideration for selection. 

For ${\cal{ P}}_1$, since the sites considered for selection at each time $t$ are only the locations in which an operational site is placed at any time $t \in T$, no information about the selection of sites has been added to the never--sampled regions in $\cal S$. Consequently, when estimating the levels of black smoke via the estimation of the latent Gaussian fields in these regions, we have no additional information about the possible values they could take. Thus, model--based estimates in these unsampled regions will tend towards the predicted global mean levels, which in this case is precisely the ${\cal{ P}}_1$--mean (the average taken across the network of observed locations). Furthermore, given the high average lifetime of the monitoring sites, estimates of the site--specific trajectories and hence the ${\cal{P}}_1$--mean barely change under the joint model due to the over--determined nature of the estimation. This is in stark contrast with ${\cal{ P}}_2$ or when a point process approach is taken. These place zero counts throughout the domain $\cal S$ and hence add additional information into the never--sampled and hence under--determined regions. The lack of change in estimates of the ${\cal{P}}_1$--mean is not a problem with the model. The quadratic model showed good model fit and we therefore see the inability of the model to change the longitudinal trajectories at the observed site locations for this dataset as proof of the model's robustness -- we would almost certainly be concerned if the estimates changed dramatically at the site locations.  

If instead, when forming our predictions of black smoke at these never--sampled locations, the model had the additional information that no site was selected here at this time (i.e. $R_{i,j} = 0$ at site $\textbf{s}_i \in {\cal S} \setminus {\cal{ P}}_1$), then this would provide the model with additional information about the likely values of black smoke at this location. For example, if preferential sampling were detected by the model, such that locations in regions with above average black smoke were estimated to have a site with higher probability (i.e if $d_\beta  >$ 0), then knowledge that a site was not placed at a given location would provide (albeit only slight) evidence for the model that the black smoke level here is below the operational network average. Suppose instead that we have a whole region such as the Highlands, with no monitoring sites present at any time. Estimates of black smoke across this region could then be considerably below the average of the predicted levels at the observed site locations throughout time, depending upon the magnitude of PS detected. This idea of filling the region with zeros to indicate non--selection is the basis of the paper of \cite{diggle2010geostatistical}, the approach taken in Implementation 3, and that seen in \citet{Watson3}.

For datasets where the average lifetimes of the monitoring sites are shorter, the measurement error is higher, and/or the functional form of the temporal trend is of higher order, then this joint model framework would have a greater capacity to change estimates of site--specific trajectories, the ${\cal{ P}}_1$--mean and hence predictions throughout $\cal{S}$. This was seen in our simulation study. However, for many applications involving data collected from static monitors, little will change in inferences under a joint model with population ${\cal{P}}_1$. An example of where large differences may be witnessed is for data collected over time from mobile monitors whose location changes at each time step. In this setting we would have a very sparse data setup, with only a single observation of the process' trajectory obtained at each location. The large under--determined missing--data problem here would present the perfect opportunity to assess the ability of the joint model framework to adjust the inference.

After the extensive network redesign in 1982, the autoregressive $\beta_1^\star(t)$ process captured a sharp decline in the average logit for site selection in 1982 (see Fig 10). This process may be reflecting, among other things, the year--by--year changes in public and political moods towards pollution monitoring. The 95\% posterior credible intervals do not cover 0 and thus the drop of over half of the network in 1982 appears to be a significant event in the lifetime of the network.

Turning our attention now to the estimated parameters of the site selection process $R_{i,j}$, no clear repulsion effect $\alpha_{rep}$ was detected ($\alpha_{rep}$ = 0.08 95\% CI (-0.14, 0.31)). This implies that any clustering or repulsion effects witnessed in the data with respect to ${\cal{P}}_1$ can be attributed to the levels of black smoke alone. On the contrary, the retention effect was found to be very large 6.18 (95\% CI (6.07, 6.29)), in agreement with common sense. This finding indicates that there is a clear incentive (possibly financial) for site--selectors to maintain sites in their current locations instead of relocating them each year.

In summary, for this dataset Implementation 2 does not lead to changes in site--specific trajectories, nor does it lead to changes in estimated BS levels in unsampled regions of GB. However, we do still gain some useful insights. We find that the site--selection was in fact preferentially made (i.e. response--biased), and that the extent of this PS could not be attributed to chance alone. Furthermore, we were able to investigate the impact of other factors, such as retention effects and changing political affinities for the network expansion on the evolving operational network $S_t$. We have presented future applications where the results from implementations 1 and 2 may not agree so closely.

\subsection{Implementation 3 -- ${\cal{ P}}_2$}
\begin{figure}
\begin{center}
\includegraphics[trim={0 0 3.5cm 0.7cm},clip,scale=0.4]{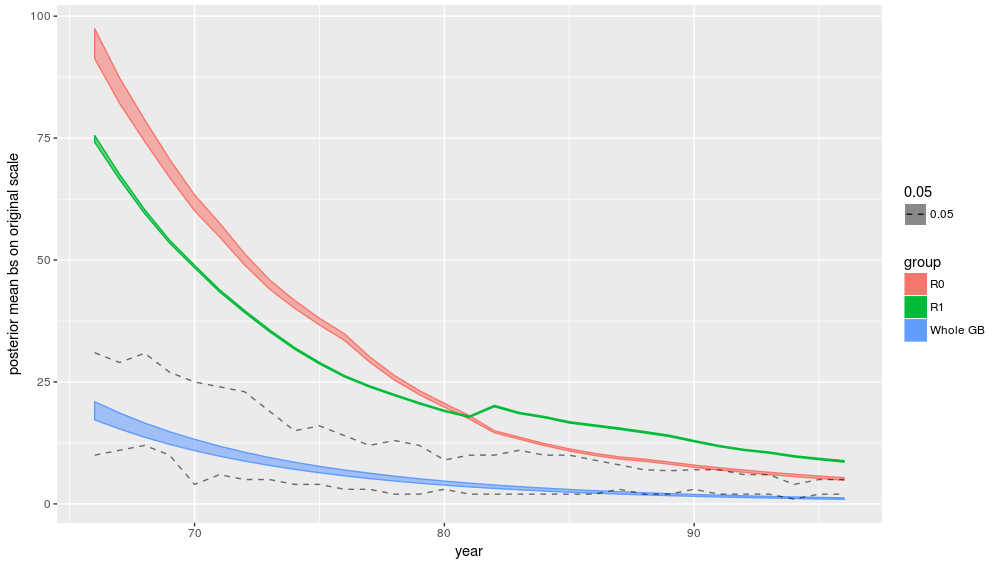}
\caption{Implementation 3. In green are the BS levels averaged over sites that were selected in ${\cal{P}}_1$ (i.e. operational) at time t. In contrast, those in red are the BS levels averaged over sites that were not selected in ${\cal{P}}_1$ (i.e. offline) at time t. Finally, in blue are the BS levels averaged across Great Britain. Also included with the posterior mean values are their 95\% posterior credible intervals. The black dashed lines denote the lower 10th percentile and lower quartile observed in the data. Note that the estimated black smoke trajectories from the pseudo--sites are \textbf{not} included in the mean calculations to form the red band. If printed in black-and-white, the green band is initially the \textbf{middle} line, the red band is initially the \textbf{upper} line and the blue band is initially the \textbf{bottom} line.}
\label{fig:Post_Mean_sites_imp3}
\end{center}
\end{figure}

Firstly, we consider the posterior parameter estimates for the two sources of PS (see Table 1). These are denoted by $d_\beta, d_b$, the medium-range and short-range preferentiabilities respectively. The posterior estimated value of $d_\beta$ was 2.77 with a 95\% posterior credible interval (2.76, 2.79). The posterior estimate of the short-range preferentiality was 0.12 with a 95\% posterior credible interval (0.11, 0.13). Thus in both cases the direction of preferentiality was significantly positive, suggesting that year-by-year, the site placements were positively associated with the relative levels of black smoke at the site location, both locally and regionally. 

Fig \ref{fig:joint2_map_plots}$^\star$ shows a striking difference in the appearance of the estimated black smoke field through time. A direct consequence of the strong preferential sampling detected is the dramatic drop in the posterior predictions of black smoke levels in undersampled regions of GB relative to Implementation 1. Fig \ref{fig:joint2_map_plots}$^\star$ shows a huge drop in estimated levels in the unsampled regions of Northern Scotland, Mid Wales and the foot of Cornwall relative to Fig \ref{fig:naive_map_plots}$^\star$ and Fig \ref{fig:joint1_map_plots}$^\star$. Implementations 1 and 2 estimated these regions to have average BS levels due to the lack of any additional information in these regions. Furthermore, Fig \ref{fig:Post_Mean_sites_imp3} shows that the posterior mean black smoke level averaged across GB is around a quarter of the size of that estimated from implementations 1 and 2 (see Fig \ref{fig:Post_Mean_Sites_Plot_Imp1} and Fig \ref{fig:post_mean_sites_plots_imp2}$^\star$). This is a direct consequence of the decreased levels estimated in the undersampled regions that make up a large percentage of the surface area of GB. This addresses objective 3 of the analysis. 

Interestingly, model inferred black smoke levels in these unsampled regions have very high standard errors (i.e. large pointwise posterior standard deviations) associated with their point estimates. This can be seen in the bottom two plots of Fig \ref{fig:joint2_map_plots}$^\star$. Here, the upper 95\% pointwise credible intervals actually cover the estimates from Implementation 1. As expected, the posterior estimates of the observed site trajectories (both operational and offline) change very little (see Fig \ref{fig:Post_Mean_sites_imp3}).  

To address Objective 4 refer to Fig \ref{fig:Post_Mean_sites_imp3}. In agreement with Figures \ref{fig:Post_Mean_Sites_Plot_Imp1} and \ref{fig:post_mean_sites_plots_imp2}$^\star$, it appears that the magnitude of preferentiality increases over time. Initially, the annual averages at the locations of the offline observed sites far exceed those from the locations of the operational observed sites. The difference diminishes over time until the major network redesign in 1982, which led to a change in direction of the relative annual mean levels. Thus it appears that the magnitude of the bias in the reported annual black smoke levels from the operational network, relative to the Great British average increased over time - with a dramatic step-change seen in 1982. Of most importance however is the discovery that the observed black smoke levels from the network appears to have \textbf{never} been representative of the levels of GB as a whole, with a positive PS effect detected at all times. In fact, Fig \ref{fig:joint2_map_plots}$^\star$ shows that around 85-90\% of the sites in the network were placed in regions with above ${\cal{P}}_2$--mean BS throughout the lifetime of the network.

Once again the autoregressive $\beta_1^\star(t)$ process reflecting the year-by-year changes in public and political mood towards pollution monitoring, captured a sharp decline in the average log intensity for site placement in 1982. The estimate is almost identical to that seen in Implementation 2 (see Fig \ref{fig:AR1_plot_imp2}) and so we omit the plot.

Regarding the estimated parameters of the site selection process $R_{i,j}$, the $\alpha_{rep}$ term was detected to be \textbf{positive} with value 0.82 [95\% CI (0.62, 1.02)]. This implies that there is additional clustering present that cannot be explained due to the levels of black smoke alone. This may be capturing some of the latent factors influencing the selection of monitoring sites such as population density. 

\subsection{Impacts of preferential sampling on estimates of population exposure levels and noncompliance}

Whilst the dramatic decline in GB--average black smoke levels seen under the joint model in Implementation 3 is interesting, the monitoring network was not intended for the accurate mapping of black smoke across the whole of Great Britain but instead was established for tracking the progress achieved by the Clean Air Act in reducing the exposure levels of both black smoke and sulphur dioxide \citep{McMillanUnpublishedAirQualityHealthEffects}. Thus judging the monitoring network based on its ability to represent the levels of black smoke across GB as a whole is potentially misleading. Taking this into consideration, we now attempt to assess the effects of PS on estimates of population exposure, and hence the effects of PS on the ability of the network to fulfill its objectives. Over the time period of study, various EU limits and guidelines on annual black smoke levels were introduced, including the annual average guide value of 34$\mu$gm$^{-3}$ introduced in 1980 (repealed in 2005) \citep{zidek2014unbiasing}. We repeat the analysis of \citet{zidek2014unbiasing} and assess the changes in the estimates of noncompliance under PS.

For estimating the population exposure levels, we obtained gridded residential human population count data with a spatial resolution of 1 km x 1 km for Great Britain based on 2011 Census data and 2015 Land Cover Map data from the Natural Environment Research Council Centre for Ecology \& Hydrology \citep{UKgriddedpopdata}. The data came in the form of a raster layer and we formulate our estimate of population density across the time period (1966 - 1996) by normalizing the count raster by dividing each cell by the total sum across all the cells. Here we assume that the relative population density has remained stable from 1966-2011 for the estimated population density layer to be a good proxy across the years of study. We also assume that residential population density is a good proxy of where the population is situated throughout the year and hence that actual black smoke exposure levels are similar to estimated residential levels. Next, we define a projector matrix, to project the GMRF estimated in INLA on the triangulation mesh onto the centroids of the population density cells that make up the raster.

Finally, we are able to use the Monte Carlo samples from the posterior marginals from INLA and the projector matrix to estimate the posterior distribution of the black smoke field at each of the grid cells. Letting $\rho_j(\textbf{s})$ denote the population density of Great Britian at location $\textbf{s} \in {\cal{S}}$, in year $j$, such that $\int_{\cal{S}} \rho_j(\textbf{s}) \textnormal{d}\textbf{s} = 1$, we can then estimate the population--mean exposure levels by approximating the following integral:

\begin{align*}
    \mu_{pop, j}(\cal S) &= \int_{\cal S} \mu(\textbf{s}, j) \rho_j(\textbf{s}) \textnormal{d} \textbf{s} \\
    &\approx \sum_{i=1}^G \Bar{\hat{\mu}}_j(\textbf{s}_i) \hat{\rho}_i = \frac{1}{M} \sum_{m=1}^{M} \sum_{i=1}^G \hat{\mu}_{i,j,m}(\textbf{s}_i) \hat{\rho}_i
\end{align*}

\noindent where $\textbf{s}_i$ denotes the $i^{th}$ raster grid cell centroid ($i = 1,...,G$), $ \Bar{\hat{\mu}}_j(\textbf{s}_i)$ denotes the Monte Carlo mean black smoke level at location $\textbf{s}_i$ in year $j$ and $ \hat{\rho}_i$ denotes the estimated population density at the $i^{th}$ grid cell. Approximate credible intervals for this quantity can also be formed. We can also use this method to estimate the proportion of the population exposed to annual average black smoke levels exceeding the EU guide level of 34$\mu$gm$^{-3}$ each year, by simply replacing the term $ \hat{\mu}_{i,j,m}(\textbf{s}_i)$ in the summation by the indicator variable representing the event that the value exceeds 34$\mu$gm$^{-3}$. Note here that the index $m$ denotes the Monte Carlo sample number.

We now do this, both for the estimated black smoke levels under Implementation 2 (i.e. Population 1) and again under Implementation 3 (i.e. Population 2). Note that the results under Implementation 1 are almost identical to those from Implementation 2 so we omit them in the plots.

\begin{figure}
\begin{center}
\includegraphics[scale=0.55]{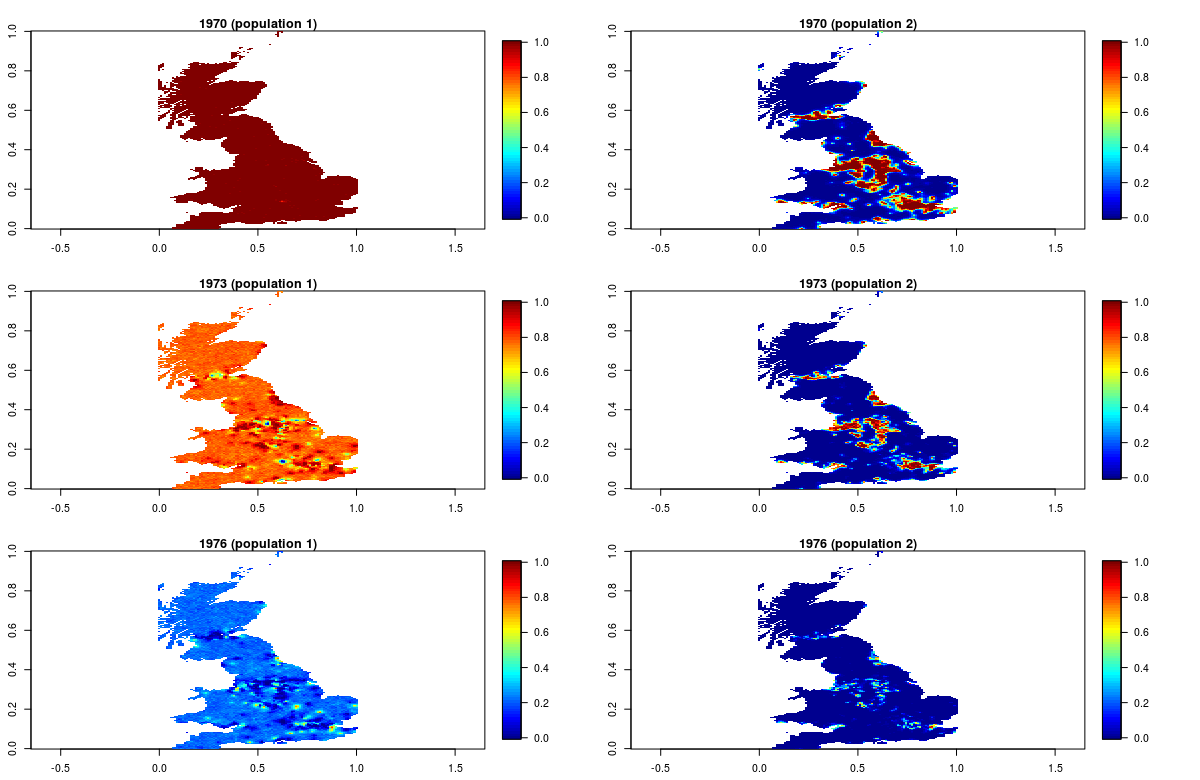}
\caption{A map plot of the posterior pointwise probability of the annual average black smoke level exceeding the EU guide value of 34$\mu$gm$^{-3}$ under Implementation 2 (left) and Implementation 3 (on the right). From top to bottom are the years 1970, 1973 and 1976. The colour scale goes from 0 to 1 for all the plots, with dark blue denoting a posterior probability of 0 and dark red denoting a posterior probability of 1. Note that the plots for Implementation 1 are almost identical to those from Implementation 2 and are omitted.}
\label{fig:exceedance_maps}
\end{center}
\end{figure}

\begin{center}
    \begin{figure}
        \centering
        \includegraphics[scale=0.5]{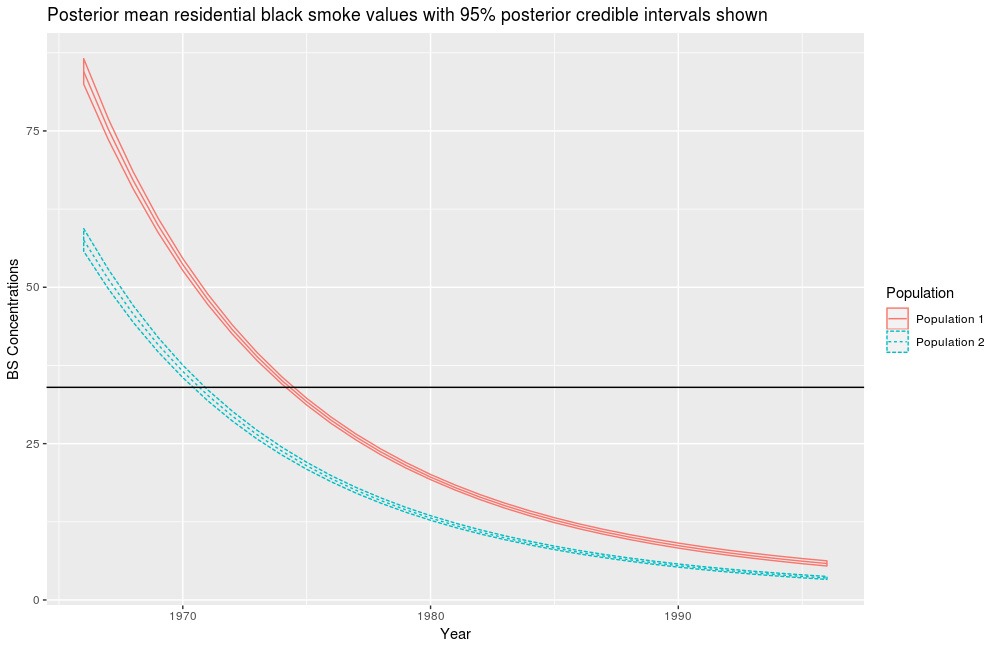}
        \caption{A plot showing the posterior mean and 95\% credible intervals of the annual residential--average exposure levels across the years of study. Shown are the results from Implementation 2 (i.e. Population 1) and from Implementation 3 (i.e. Population 2). The horizontal line denotes the EU guide value for annual average black smoke levels of 34$\mu$gm$^{-3}$.}
        \label{fig:residentialexposuremean}
                \includegraphics[scale=0.5]{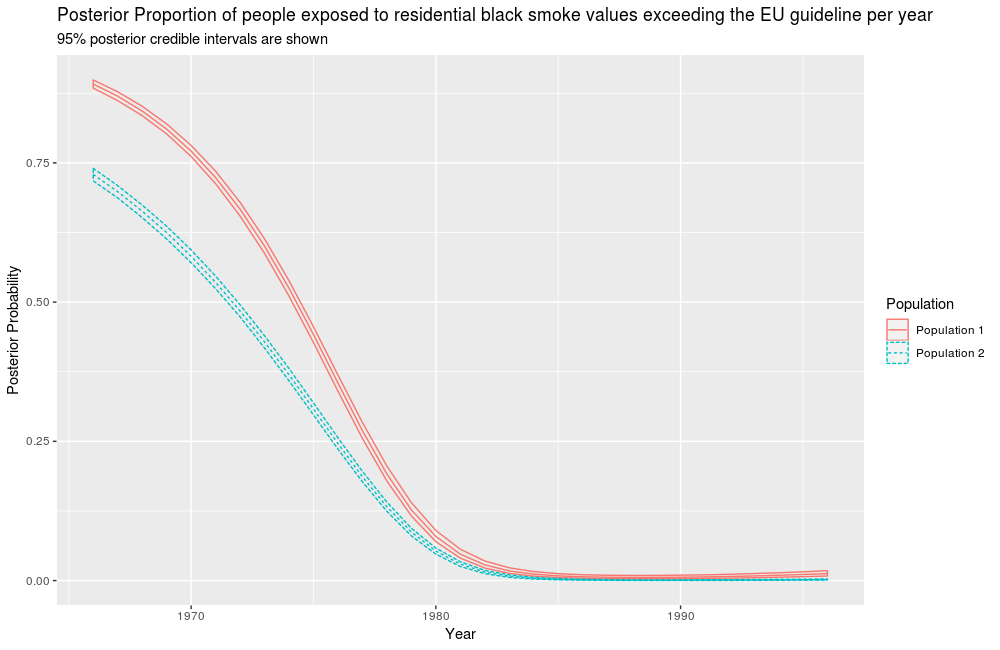}
        \caption{A plot showing the posterior mean and 95\% credible intervals of the annual proportion of the population with black smoke exposure levels exceeding the EU guide value of 34$\mu$gm$^{-3}$ across the years of study. Shown are the results from Implementation 2 (i.e. Population 1) and from Implementation 3 (i.e. Population 2).}
        \label{fig:proppopexceedance}
    \end{figure}
\end{center}

Fig \ref{fig:exceedance_maps} shows plots of the posterior pointwise probability of exceeding the EU annual black smoke guide value of 34$\mu$gm$^{-3}$ under implementations 2 and 3, across the years 1970, 1973 and 1976. The colour scale goes from 0 to 1 for all the plots, with dark blue denoting a posterior probability of 0 and dark red denoting a posterior probability of 1. In agreement with the plots of the pointwise posterior means (see Fig \ref{fig:joint1_map_plots}$^\star$ and Fig \ref{fig:joint2_map_plots}$^\star$), a dramatic decline in the estimates of noncompliance can be seen under Implementation 3 in the regions far from the nearest monitoring network across the years (see Fig \ref{fig:exceedance_maps}). This has major ramifications regarding the total reported proportion Great Britain in noncompliance with the guide value. For example, under Implementation 2 almost the entirety of Great Britain is estimated to be in noncompliance with the guide value up until 1970. This figure drops to below 25\% in 1970 under Implementation 3 (see Fig \ref{fig:prop_exceedance_landarea}$^\star$). 

However, once again the monitoring network and the guide value were intended to measure and control the population exposure to black smoke levels. Thus our maps showing the pointwise posterior probability of exceedance, whilst being dramatic, may not be a fair assessment of the network. Instead, we now focus our estimates on the estimated proportion of the population of Great Britian exposed to black smoke levels out of compliance with the air quality standard. Given that the density of monitoring sites in the network follows the large population centres of GB closely, we expect the differences between the estimates to be much lower. In fact, this is not the case. Fig \ref{fig:proppopexceedance} still shows a large decrease in the estimated proportion under Implementation 2, from 89\% to 73\% in 1966 for example. Note that the posterior credible intervals still show a large discrepancy between the estimated proportions. This is despite us including the additional short scale variability from the spatially-uncorrelated IID effects in the estimates (one pair of realised $b$ terms per 1km grid cell, per Monte Carlo sample). 

Finally, we turn our attention to the estimated population--average annual black smoke exposure levels across the two implementations (2 and 3). In agreement with Fig \ref{fig:proppopexceedance}, Fig \ref{fig:residentialexposuremean} shows a clear decrease in the estimated annual averages. Given the sensitivity of health effect estimates of air pollution to the accuracy of population exposure levels, this result is especially striking.

\section{Discussion}
 
Importantly, a lot of the detected preferentiality effects and subsequent de--biasing effects on prediction are likely mediated by well--known covariates. For example, annual population density figures and/or industrialisation indices (in their correct functional form) would likely simultaneously explain a lot of the PS detected if included in the $R_{i,j}$ process, and be strongly positively associated with the observed levels of $Y$ in the observation process. Sites may well be placed in regions where lots of people live and work to ensure the network captures `typical' exposures experienced by the public, and some sites may be located in areas close to polluting industry for exceedance detection. Since the daily activities of people and industry may well be the main contributors to black smoke levels, including these covariates in the observation model $Y$ would therefore likely lead to decreased model--estimated pollution levels in unsampled regions such as The Highlands of Scotland with low population density and industry.  

In many applications, the preferential sampling may disappear upon the inclusion of such covariates and hence be reduced to a missing--at--random scenario. Given that the focus of this paper was to repeat previous analyses of this dataset \citep{shaddick2014case, zidek2014unbiasing} under our new framework and assess the changes, we do not consider including covariates here. Furthermore, we wanted to show that in settings where such covariates are unavailable, sensible adjustments can still be realised under a careful use of our model framework.
Additionally, given that the locations of the monitoring sites are almost exclusively situated near population-dense, industrious and urban regions, it is unclear if these locations would provide the adequate contrast required to estimate the correct the functional forms of these covariates. It would be interesting in future work to see if any PS is detected in this data after conditioning on as many such variables in both processes. In summary, this paper is not attempting to bypass the need for including relevant covariates in the modelling. Rather, it is presenting a method for accounting for the effects of any \textbf{residual} unmeasured confounders associated with \textbf{both} processes by using spatio--temporal fields to act as a proxy.

It is the authors' view that this modelling framework should be considered to both detect preferential dropout within a fixed population or network $\cal P$, \textbf{and} to detect if the population or network $\cal P$ was preferentially placed within the domain of study $\cal S$. Accomplishment of both of the above depends upon the choice of population of sites under consideration for the site-selection process. If preferential sampling is detected using this model, then first and foremost, the modeller should attempt to find available covariates that mediate the detected preferentiality. If, after exhausting the available mediators (e.g. population density), and after removing as many sources of variability from the site--selection process as possible, preferentiality is still detected, then this modelling framework should be used for detecting the potential consequences of this sampling scheme on the subsequent inference -- either on parameters or spatio-temporal prediction. 

Furthermore, different regression models can be explored for the initial site--placement and site--retention processes. For example, different covariates may be believed to affect only one of the two processes, the qualitative behaviour of certain covariates on the two processes may be different or perhaps the nature of PS could differ across the two processes. We didn't explore these possibilities here, assuming only a unique intercept existed between the two processes. This extension is explored in \cite{Watson3}.

Additionally, the functional form used to model PS can be as flexible as desired. Here we opted to model the direction and magnitude of preferential sampling as being constant through time. In reality this may not be suitable and the direction and magnitude of preferentiality may change through time. In Fig 9 we can see that initially (at $t = 1$) the operational network was established such that it gave annual readings below the $\cal P$--mean under Population 1. Then, as time progressed, the magnitude of the preferentiality decreased as the annual averages from the operational sites approached those from the population average. Thus it may make sense here to estimate a separate preferentiality parameter $d_\beta$ for times 1 and for $t > 1$. For time 1 this would likely be estimated to be smaller compared with for $t > 1$. For simplicity we opted against this approach, however such a model would help paint a more detailed picture of the dynamic nature of the PS through time.  

If one wishes to adjust the estimates of the domain--average (the GB--average in our example) to the effects of PS, the population of locations $\cal P$ considered for selection should be extended to include locations in unsampled regions in the domain of study $\cal S$. Population 2 did just that, and as a result the GB--average estimates significantly dropped under the joint model. An alternative approach would be to consider modelling the site placement events each year implicitly as realisations from a LGCP and the site retention events separately as Bernoulli trials \citep{Watson3}. Two reasons for not pursuing this approach were given earlier in the paper. An example where extending the population of locations for selection $\cal{P}$ beyond the network locations would  be the case when we knew with certainty that the monitoring sites within the network were placed in $\cal{S}$ independently from the environmental process under study and also to any covariates or processes that may have been associated with it. An example of this is if the monitoring sites were located in $\cal{S}$ completely at random. 

Extensive analytic and simulation studies on jointly modelling dropout with various longitudinal clinical markers have been made in biostatistics over the past 20 years. The authors' of this paper gained their inspiration for this work from the literature on the joint modelling of viral load, dropout and longitudinal clinical markers measured in HIV clinical trials \citep{wu2009mixed,li2018joint,lawrence2015joint}. In fact, after transforming the data, Fig \ref{fig:Spaghetti_plot} shows black smoke trajectories that are very similar to the subject--specific dose--response trajectories seen in such longitudinal clinical data. The same philosophy behind jointly modelling informative patient dropout (i.e when the dropout violates the missing at random assumptions) with the process of interest via shared random effects can be applied to spatio--temporal environmental network data with minimal alteration. The major difference with spatio-temporal data are the spatial correlations assumed on the random effects. It is this correlation which allows for the spatial extrapolation to occur. 

Whilst the case study in this paper considered the observations to be on the same time scale as the site--selections, this need not be the case. For example, this general framework could simultaneously model high--frequency (e.g. hourly) observations with a low--frequency (e.g. annual) site--selection process. This would comprise decomposing the temporal trajectories into trend, seasonal and cyclical (e.g. daily) terms in the model. It would then likely make most sense to include only the trend term in the linear predictor of the site--selection process.

Assuming the locations of the monitoring sites are realisations from an inhomogeneous Poisson point process (IPPP) or a LGCP, while being useful computationally, may not always be sensible in certain applications. For example, if a strict lower limit on the distances between the monitoring site locations was known, then a LGCP or a IPPP would not be the most suitable model for use and alternatives such as a Matérn hard-core point process model would be more suited \citep{baddeley2015spatial}. Having said that, a nice property of using our logistic regression approximation to the LGCP, is that we are able to delete pseudo--site locations in ${\cal{P}}_2$ that violate any known rule (e.g. a minimum distance/hard--core rule). Furthermore, if additional clustering is present then a cluster point process or Gibbs point process may be more desirable \citep{baddeley2015spatial}. Whilst we attempt to adjust for the additional clustering seen in our dataset by constructing a covariate $I_{i,j}$, this is by no means the best way forward here.   

On a closing note, it should be apparent that the modelling framework introduced in this paper can be applied to monitoring data that have come from static monitoring sites, mobile monitoring sites, and a combination of the two. Furthermore, the ability for the joint model framework to adjust for PS under ${\cal{ P}}_1$ should be greater in applications with mobile monitors. One such study that could be revisited is the MESA Air Study (\url{http://www.mesa-nhlbi.org/}). Since this study involves the estimation of the health effects associated with exposure to various air pollutants, with pollution readings taken from a combination of static and mobile monitoring sites, this data set offers an ideal opportunity to test out this framework. Of interest may be the detection of any preferential sampling, and its resulting effects on the health effects.

\section{Conclusion}

We applied our general framework to the network of air quality monitors in Great Britain between the years 1966-1996. From this, we were able to show that the monitors were preferentially placed within Great Britain throughout the life of the network. In particular, each year the locations of the operational sites were found to have been situated in areas with black smoke levels considered much higher than the annual average level across Great Britain. Furthermore, we showed that the network was updated in a preferential manner throughout the life of the network. Monitoring sites at locations with highest black smoke levels were favoured for selection into the network each year, and monitoring sites at locations with lowest black smoke levels were favoured for removal from the network each year. 

The implications for this biased network placement were then clearly demonstrated. The preferential sampling of the monitoring sites may have had a significant deleterious impact upon the ability of the network to serve its purpose as a tool for measuring the black smoke exposure levels experienced by the population of Great Britain as a whole. It appears that estimates of population exposure levels may have been overestimated (see Fig \ref{fig:residentialexposuremean}). Furthermore, estimates of noncompliance to the various air quality regulations established throughout the chosen time period of 1966 - 1996, may also have been affected by how and where the monitoring sites were situated. It appears that any estimates of noncompliance that used the observations from the air quality monitoring network may have over--estimated the true amount of noncompliance (see Figures \ref{fig:exceedance_maps} and \ref{fig:proppopexceedance}). This includes historical estimates of the proportion of the population of Great Britain exposed to black smoke levels that were out of compliance.

\section{Acknowledgements}
The authors would like to thank the Associate Editor and the anonymous referees for their insightful comments. Their constructive feedback greatly helped to improve the focus of this paper.

\bibliographystyle{plainnat}
\bibliography{prefsamp.Jan03_2017,prefsamp.bib,BookRefs_20150106_JZ,personal_work.bib,bibliography.bib,bibliography_Jim_rev.bib}

\newpage

\section{Supplementary Material}

\subsection{Chosen priors for the case study}
For the $Y$ process, we used weakly informative Gaussian priors for the $\gamma_k$'s. We used a Gamma($a$, $b$) prior for the precision parameter $1/\sigma^2_{\epsilon}$, where $a$ denotes the shape parameter and $b$ denotes the inverse-scale parameter. We chose $a = 1$ and $b = 5 \times 10^-5$. Under this parameterisation, the mean and variance of this distribution are $a/b$ and $a/b^2$ respectively. Thus this prior assumption allows for very large and very small variances of the response to exist. Next, the a 2D Wishart distribution is assumed for $\Sigma_b^{-1}$ with four degrees of freedom. The prior matrix is given 0 off--diagonal elements and diagonal values of 1. This results in a prior mean for the two variance terms of the random effects ($\sigma^2_{b,1}, \sigma^2_{b,2} $) of 4 with a prior variance for these terms equal to 8. The prior mean for the correlation term is 0 with variance for the logit transform of the correlation equal to 4. This allows for random effects with a large range of magnitudes and correlation structures to exist. We place the PC joint priors \citep{simpson2017penalising,fuglstad2017constructing} on the two hyperparameters for the 3 independent Matern realisations, with prior belief that the lower 5th percentile for the range is 3.4km (a fifth of the smallest range found in previous analyses) and the upper 1st percentile for the standard deviation of each field is 1 (noting that the data have been transformed). We fix the Matern roughness parameter to equal 1 since this is the largest smoothness value currently implemented in R-INLA, and we assume a-priori that the medium--range pollution process will be reasonably smooth. The lower prior bound on the range parameter, combined with the probabilistic upper bound on the variance, should help prevent the model from collapsing into a state that over--fits the data.

For the site--selection process $R$, our choice of priors follows the same objectives as for the observation process. Weakly informative Gaussian priors were placed on all the $\alpha$ terms. The same PC prior chosen for the observation process was placed on the $\beta_0^\star(\textbf{s})$ field. For the first order autoregressive term $\beta_1^\star(t)$, we placed a Gamma$(1, 5 \times 10^-5)$ on the marginal precision and a $N(0, 0.15)$ prior was placed on the logit of the lag 1 correlation (i.e. on $\textnormal{log}\left( (1+\rho_a)/ (1-\rho_a) \right)$) to allow for a large degree of flexibility. Finally, we consider two different sets of priors for the PS parameters $d_b, d_\beta$. For Implementation 1 we constrain these to equal 0 and thus we can view this as setting a point mass prior at 0. For implementations 2 and 3, we assign a $N(0, 10)$ prior to allow PS to be detected.

\subsection{Details on the R-INLA implementation}

We used the estimated ranges from \citet{shaddick2014case} to construct the Delauney triangulation mesh required for use in R--INLA. Following the advice of \cite{bakka_2017,mesh_test}, and trading it off with the need for maintaining a reasonable computation time, we set the edge lengths of the triangles throughout the domain to be around 5km, less than the minimum estimated range of 17km found in \citet{shaddick2014case}. This is important since it has been shown that the length of the triangle edges must be less than the range of any Matern field and should ideally be less than a quarter of this. Failure to do so leads to large errors in the approximation of the Gaussian random field. We are confident that with our choice of mesh, any changes to the inference in the unsampled regions will be a direct result of our joint model framework and not due to any undesirable artifacts caused by a poor choice of triangulation mesh for the SPDE approximation. 

It is well known that an empirical Bayes or maximum likelihood approach does not fully account for the uncertainties in the hyperparameters when performing predictions and inference, and these may be high in spatially correlated Gaussian random fields \citep{zhang2004inconsistent}. Interestingly, for this dataset we compared the fully Bayesian approach with the empirical Bayes method using R-INLA and found little difference. The posterior credible intervals for the latent effects and parameters were slightly wider under the fully Bayesian approach, however the posterior credible intervals for the predictions were almost identical. Additionally we used the empirical Bayes approach in a small simulation study with good results. Thus for computational savings we opted to consider only empirical Bayes methods.

In R-INLA, copying across a linear combination of latent processes (potentially from a different time point) requires the use of dummy variables. In particular, the idea of \cite{ruiz2012direct} is required. This simply involves creating infinite precision Gaussian variables with observed values of zero and with linear predictor set equal to the (negative) linear combination of latent processes desired, plus an infinite variance random intercept process. It is not hard to see that the values of these random intercepts equal precisely the values of the linear combination of the desired processes. This approach proved vital for fitting implementations 2 and 3.

Note that in essence, for Implementation 3 we are modelling the initial site--placement process as a LGCP, but using a Bernoulli likelihood as a pseudo-likelihood instead of the usual Poisson likelihood to form the computational approximation. We use the conditional logistic regression approach, commonly used to fit Poisson point processes, placing the zeros in a regular (not a latticed) manner throughout $\cal S$, independent from the observed site locations. In practice, we created a reasonably regular delauney triangulation mesh in R-INLA throughout $\cal S$ for our GMRF with mesh vertices placed independent from the observed site locations. Regularity was enforced through a combination of the choices of a minimum vertex length of 5km, an upper vertex length of 7km and a minimum angle of 25 degrees. We then used the created mesh vertices as our pseudo--sites.

A somewhat undesirable property of using the logistic regression approach is that the likelihood value does not converge as the number of pseudo zeros tends towards infinity. Thus, unlike the result of using the Poisson approximation to a Point Process, convergence must instead be judged with the convergence of \textbf{fixed} parameter estimates, excluding the estimate of the intercept. However, if the Poisson approximation is chosen, then it cannot be used to simultaneously model the retention process alongside the site--placement process and hence a third Bernoulli likelihood modeling the retention--process would be required. Thus in either case, there is a trade-off. Given that the computational time required to fit the model in R-INLA using the SPDE approach is affected more by the resolution of the computational mesh than by the number of observations, we can increase the density of the pseudo--sites with a reasonably small effect on the total computation time. 

Thus for fitting Implementation 3, we follow the advice given in the literature \citep{warton2010poisson,fithian2013finite}. We repeatedly re--fit the joint model on an ever-increasing density of pseudo--sites until the parameters and predictions converge. We found that all estimates, except of course the site--selection intercept, stabilised once the average distance between pseudo--sites was decreased to 5km. This supports the claim that our estimates from our model are close to those of the joint triple model with a LGCP for the site--selection process, a Bernoulli likelihood for the site--retention process, and a Gaussian process for the observation process. 

The correct placement of the zeros in the site--selection process is vital for the asymptotic convergence of the pseudo-likelihood to the LGCP. In particular, the asymptotics of the conditional logistic regression approximation used in our example with the logit link are only established when the zeros are either a realisation of a homogeneous Poisson point process, independent of the monitoring site locations \citep{baddeley2015spatial}, or when they are placed uniformly throughout the domain $\cal S$ \citep{warton2010poisson}. In either case, the density of the zeros must be uniform (at least in probability) throughout $\cal S$ for each year $j \in \{1, ... ,31\}$ and be placed independent from the observation locations. 

A direct consequence of this is that for our site--selection process, we should \textbf{not} consider for selection at time $j$ the subset of observed sites (i.e. the subset of Population 1) that are offline at year $j$ (i.e. $S_{t_j}^C$). Put differently, we should not include the $R_{i,j}$'s in Population 1 in the likelihood such that $r_{i,j}=0$. Erroneously doing so would lead to an increased density of zeros in the heavily sampled regions and thus a `preferential sample' of zeros. Similarly, for the site--retention process at time $j$, we should only consider the sites online at the previous time $j-1$. 

Putting these two processes together, the \textbf{only} zeros that should contribute to the joint Bernoulli likelihood at time $j$ are the pseudo--sites and the sites that were online at the previous time $j-1$ and were removed from the network at time $j$. In fact, we tested the sensitivity of the results to the above, re--fitting the model once by following the advice given above, and again but ignoring the advice and considering \textbf{all} the observed sites (operational and offline) for selection at each time step $j$, along with the pseudo--sites. Despite the former being more appropriate, we found no differences in estimates, but we required a higher density of pseudo--sites, and hence an increased computational cost to reduce this bias in the parameter estimates. This advice is therefore of most importance for the modelling of very large datasets where the number of unique observed site locations through time could be much higher than seen here. 

To form all of our predictions and maps, we simulated 1000 MCMC samples of all the parameters and latent effects from the fitted models. This feature is available in the R-INLA package \citep{lindgren2011explicit,rue2009approximate,rue2017bayesian}, by simply saving all the configuration settings generated by the software required to fit the model. We then formed all the site-specific trajectories by appropriately combining all latent effects and parameters in the linear predictor. We take the mean, the empirical upper 97.5\% and empirical lower 2.5\% values of the 1000 linear predictor estimates to form our credible intervals. Finally, to obtain the map of the pointwise expectations of the predictive distribution across GB, we used the MCMC samples of the latent effects and parameters (minus the IID site-specific effects) and linearly interpolated the estimated field throughout $\cal S$ on a regular lattice grid covering the the map of GB, before taking the empirical mean and standard deviation across the 1000 maps. To compute the average BS across the Whole GB, we take the mean (averaging across the pixels) of each the 1000 sampled/realised maps. Then, we take the mean, the empirical 2.5\% and the empirical 97.5\% values of these 1000 (mean) values.

\subsection{Posterior pointwise mean and pointwise standard deviation plots}

\begin{center}
    \begin{figure}[H]
        \centering
        \includegraphics[scale=0.25]{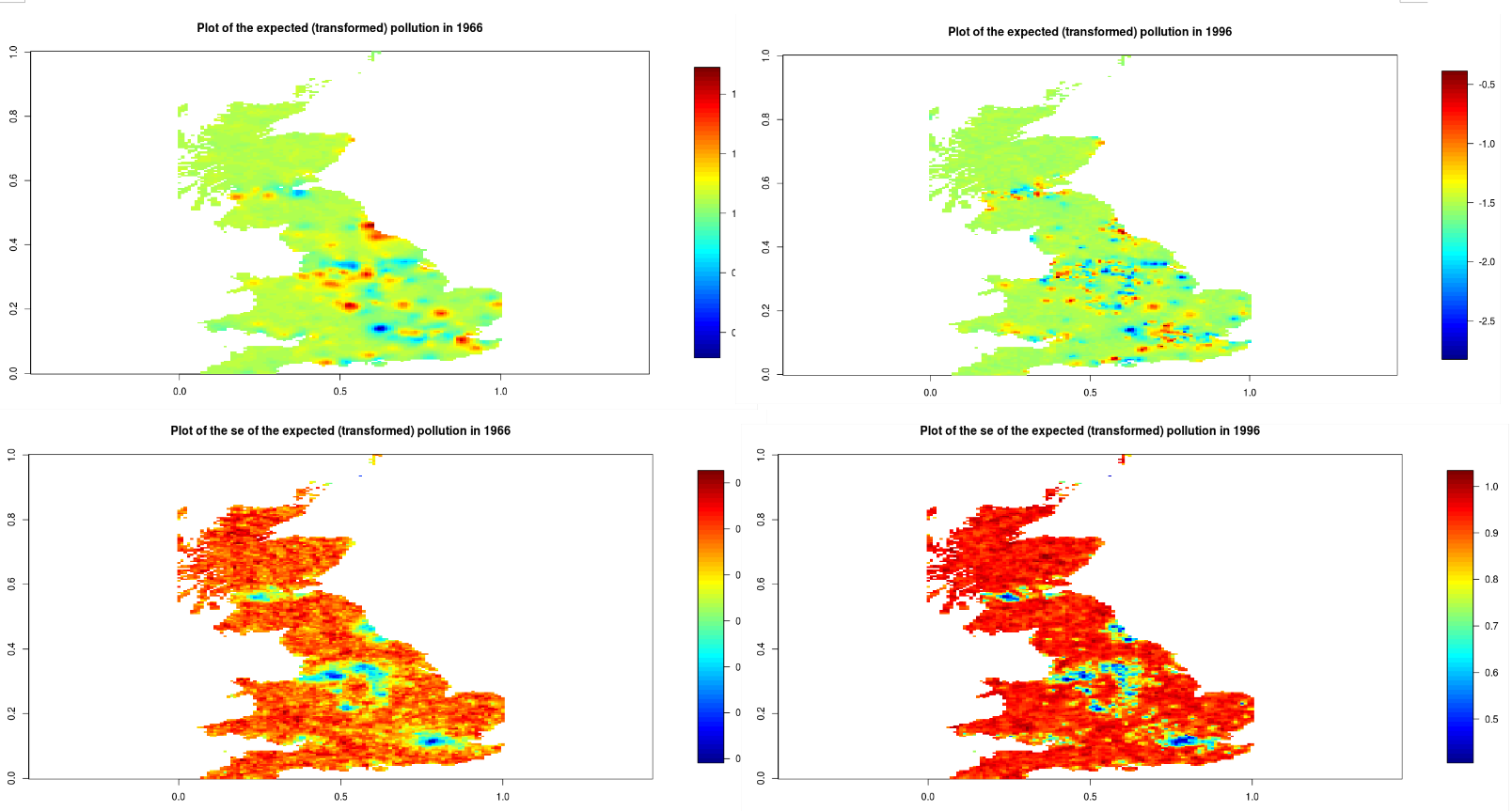}
\caption{A plot of the posterior mean black smoke in 1966 and 1996 under Implementation 1 with corresponding standard errors plotted below. Note that for visualisation purposes, the two plots have had their values scaled to put them on the same colour scale.}
        \label{fig:naive_map_plots}
    \end{figure}
\end{center}

\begin{center}
    \begin{figure}[H]
        \centering
        \includegraphics[scale=0.4]{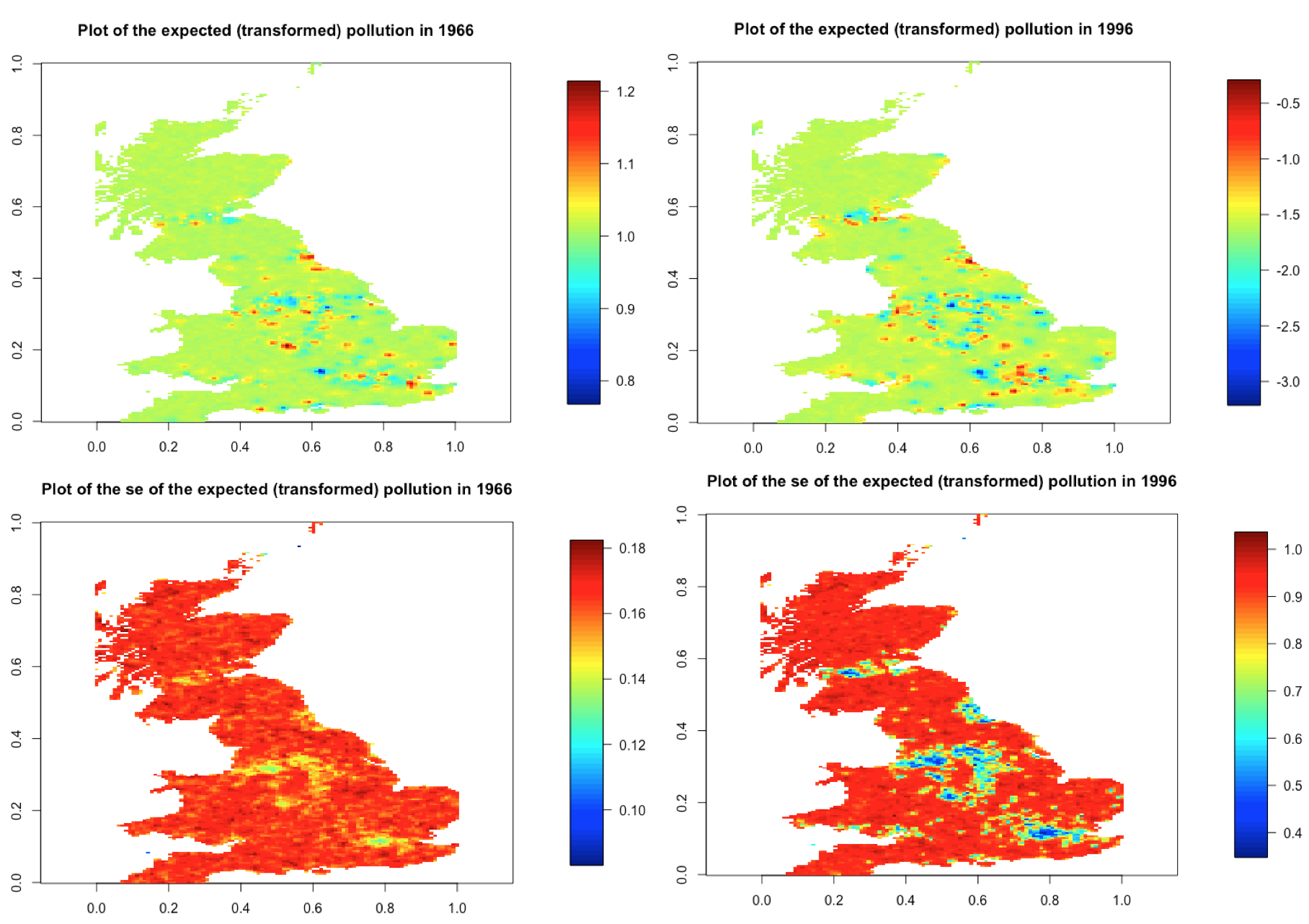}
\caption{A plot of the posterior mean black smoke in 1966 and 1996 with corresponding standard errors plotted below. Estimates are taken from Implementation 2. Note that for visualisation purposes, the two plots have had their values scaled to put them on the same colour scale.}
        \label{fig:joint1_map_plots}
    \end{figure}
\end{center}

\begin{center}
    \begin{figure}[H]
        \centering
\includegraphics[scale=0.34]{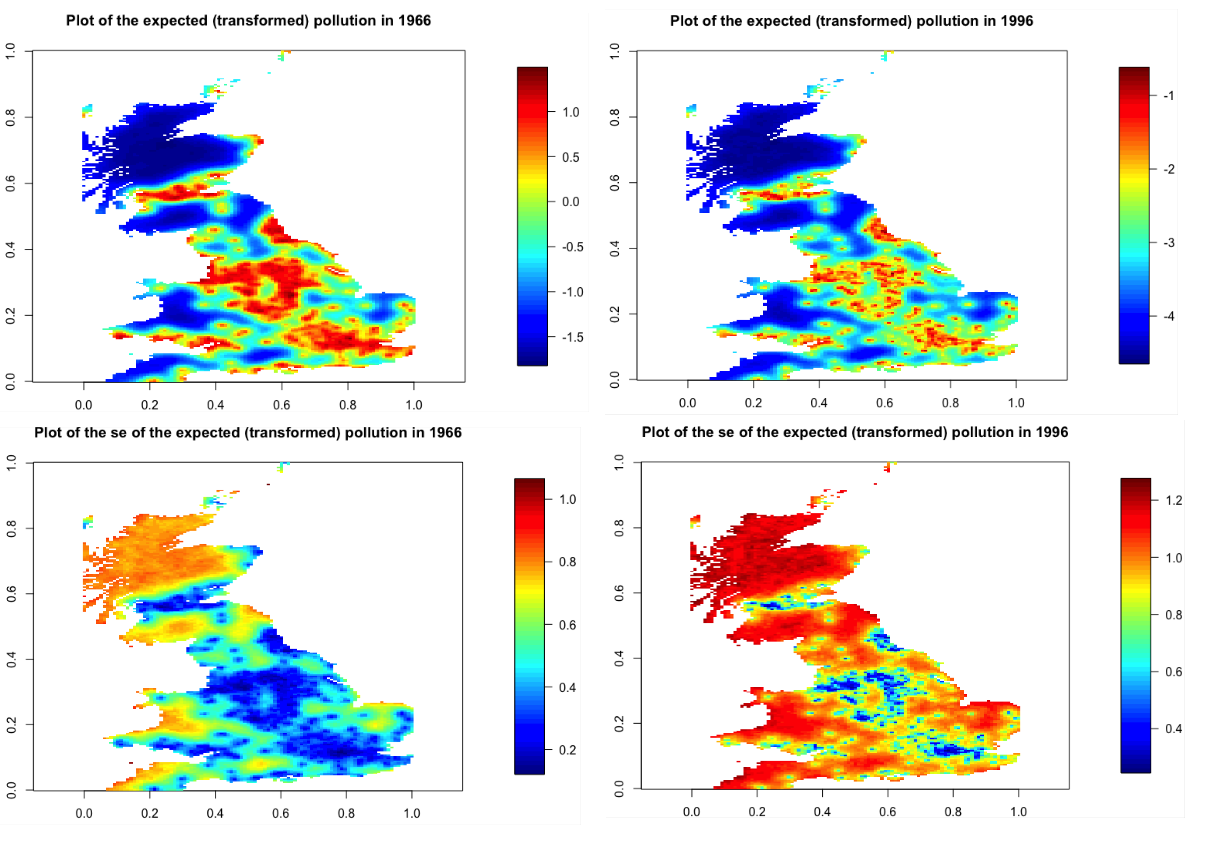}
\caption{A plot of the posterior mean black smoke in 1966 and 1996 with corresponding standard errors plotted below. Estimates are taken from Implementation 3. Note that for visualisation purposes, the two plots have had their values scaled to put them on the same colour scale.}
        \label{fig:joint2_map_plots}
    \end{figure}
\end{center}

\subsection{Additional plot of the exceedance of the annual black smoke EU guide value}

\begin{figure}[!h]
\begin{center}
\includegraphics[trim={0cm 0cm 0cm 1.3cm},clip,scale=0.5]{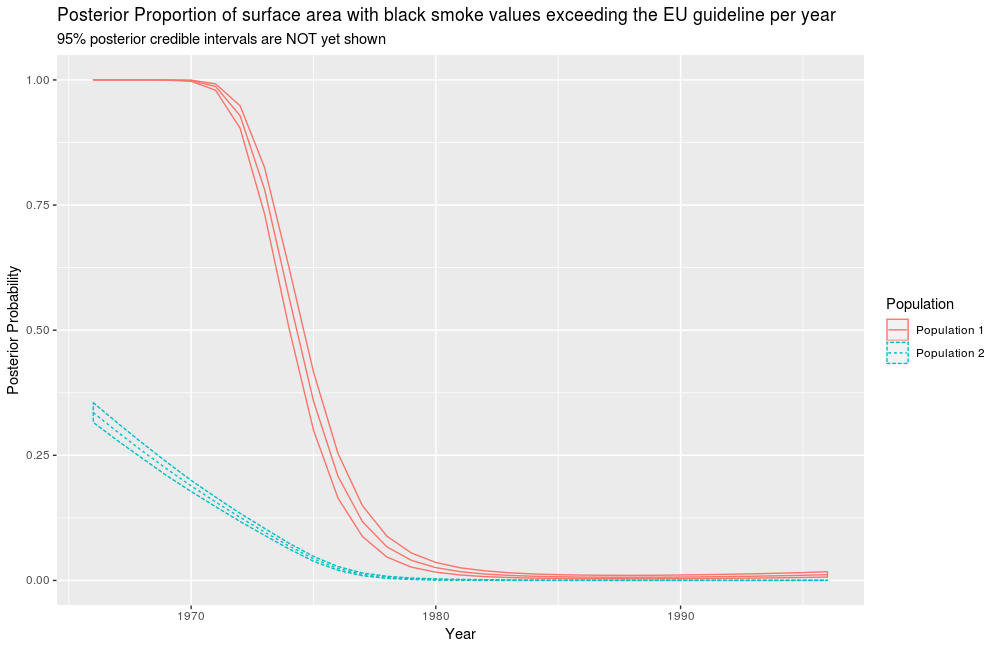}
\caption{A plot showing the posterior proportion of the total surface area of Great Britain with annual average black smoke level exceeding the EU guide value of 34$\mu$gm$^{-3}$. Shown are the results from Implementation 2 (the red solid line) and Implementation 3 (the blue dashed line). Note that the line for Implementation 1 is almost identical to that from Implementation 1 and omitted.}
\label{fig:prop_exceedance_landarea}
\end{center}
\end{figure}

\FloatBarrier

\goodbreak

\subsection{Additional plot of annual average black smoke levels}

\begin{figure}[!h]
\begin{center}
\includegraphics[trim={0 0 3.5cm 0.8cm},clip,scale=0.4]{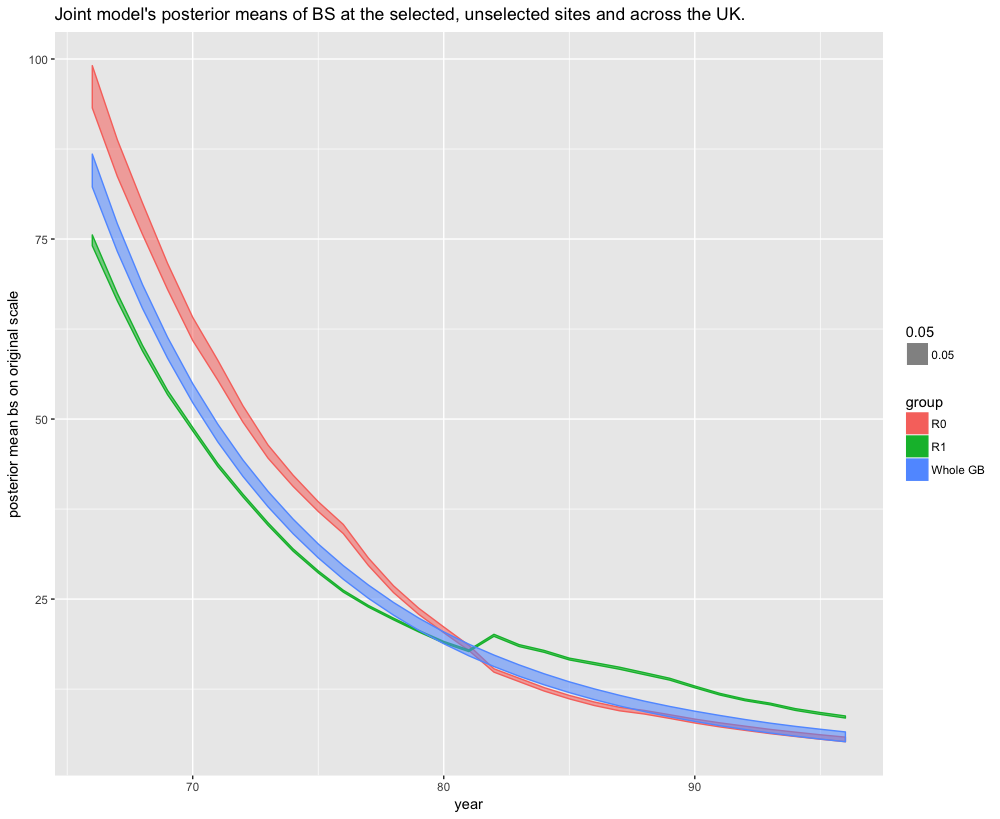}
\caption{Implementation 2. In green are the BS levels averaged over sites that were selected in ${\cal{P}}_1$ (i.e. operational) at time t. In contrast, those in red are the BS levels averaged over sites that were not selected in ${\cal{P}}_1$ (i.e. offline) at time t. Finally, in blue are the BS levels averaged across Great Britain. Also included with the posterior mean values are their 95\% posterior credible intervals. If printed in black-and-white, the green band is initially the lower line, the red band is the upper line and the blue band is initially the middle line.}
\label{fig:post_mean_sites_plots_imp2}
\end{center}
\end{figure}

\newpage

\subsection{Model diagnostic plots}

We include, for each of the three implementations considered in this paper, residual plots to help diagnose poor model fit. Included are residuals vs. year plots, with a fitted lowess smoother to help show that the choice of a quadratic model adequately captured the temporal trend in the data. Also shown are normal QQ-plots of the residuals with fitted 99\% confidence bands around the overlain QQ-line. It is clear from this plot that a heavier tailed distribution on the response would have been more suitable. Finally, we include histograms and normal QQ-plots of the random effects. Here we see slightly left-skewed and right-skewed empirical marginal distributions for the random intercepts and slopes respectively, however strictly speaking we should consider the empirical joint distribution of these effects. We have no strong cause for concern with these final plots.  

\begin{figure}[H]
\begin{center}
\includegraphics[scale=0.5]{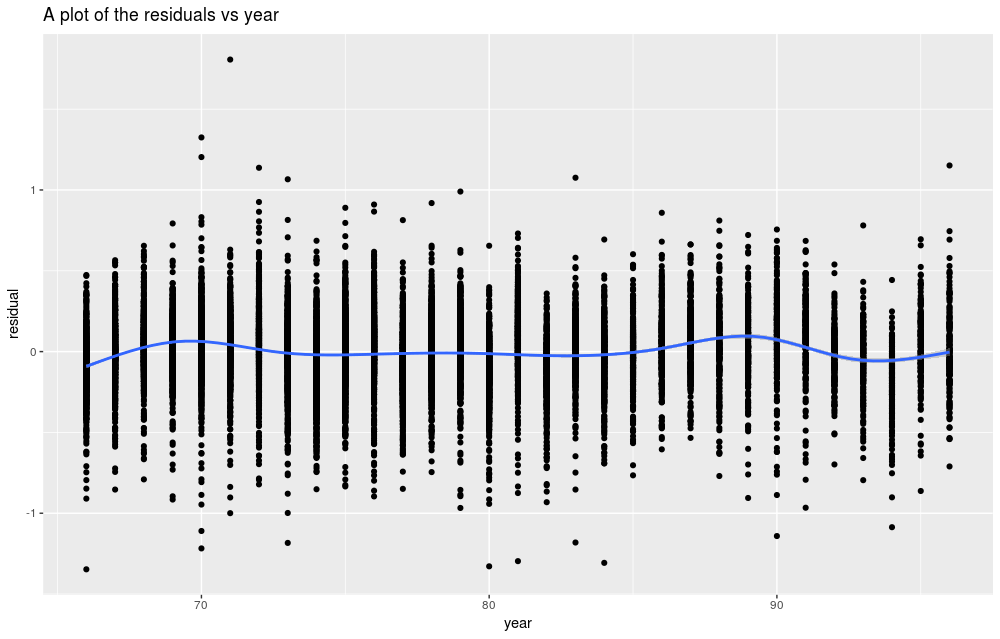}
\caption{A plot of the residuals vs. year from Implementation 1 with a fitted smoother.}
\label{fig:naive_residuals}
\includegraphics[scale=0.7]{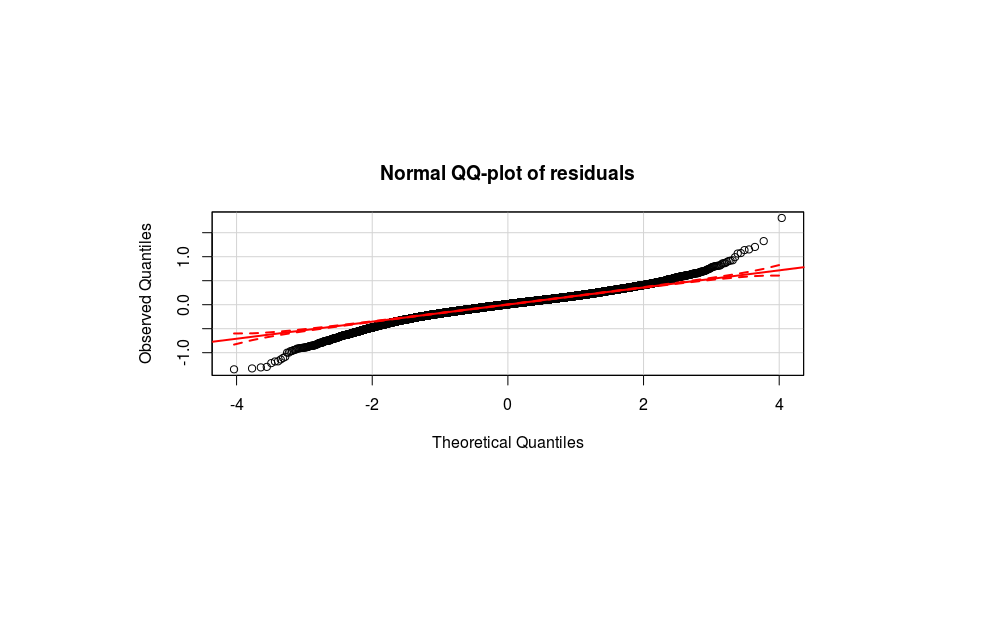}
\caption{A Normal Q--Q plot of the residuals from Implementation 1.}
\label{fig:naive_QQ}
\end{center}
\end{figure}

\begin{figure}[H]
\begin{center}
\includegraphics[scale=0.7]{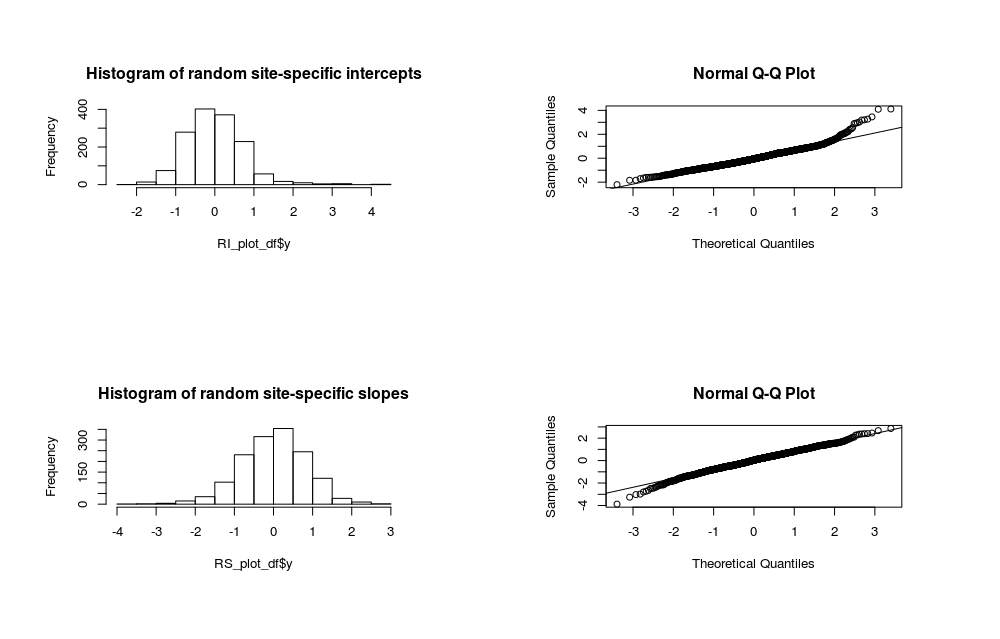}
\caption{Histograms of the spatially--uncorrelated random intercepts (top left) and slopes(bottom left), with corresponding Normal Q--Q plots shown on the right from Implementation 1.}
\label{fig:Naive_RI_plots}
\end{center}
\end{figure}

\begin{figure}[H]
\begin{center}
\includegraphics[scale=0.3]{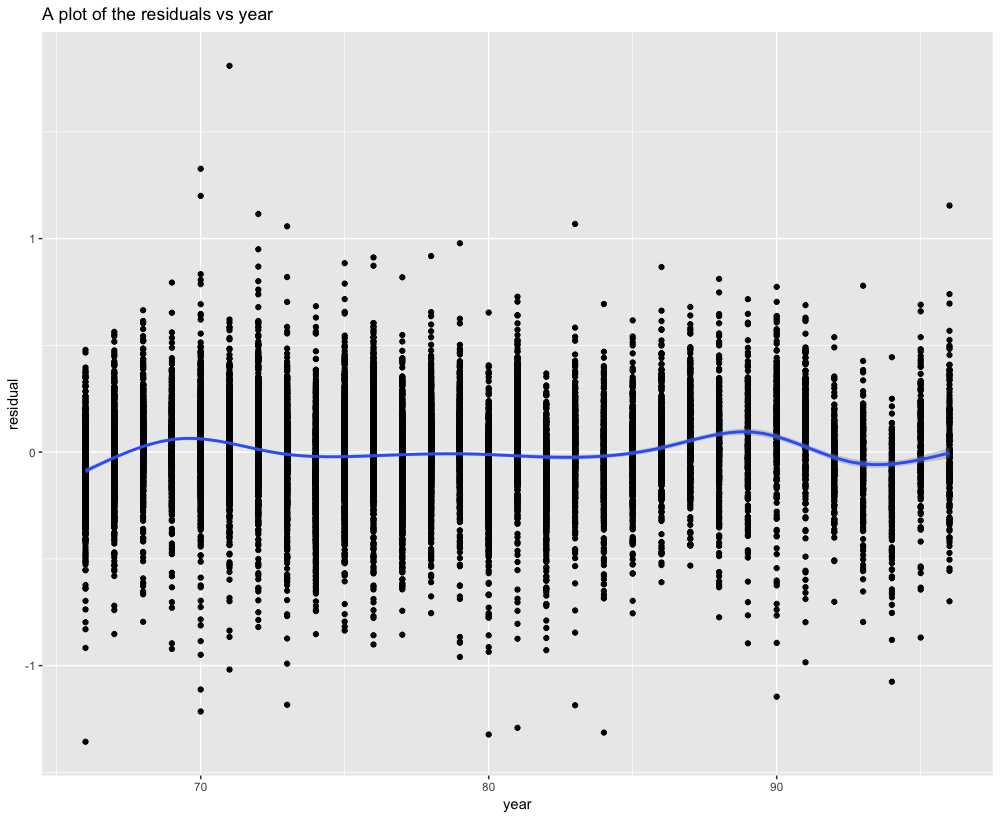}
\caption{A plot of the residuals vs. year for Implementation 2, with a fitted smoother.}
\label{fig:joint1_residuals}
\includegraphics[scale=0.4]{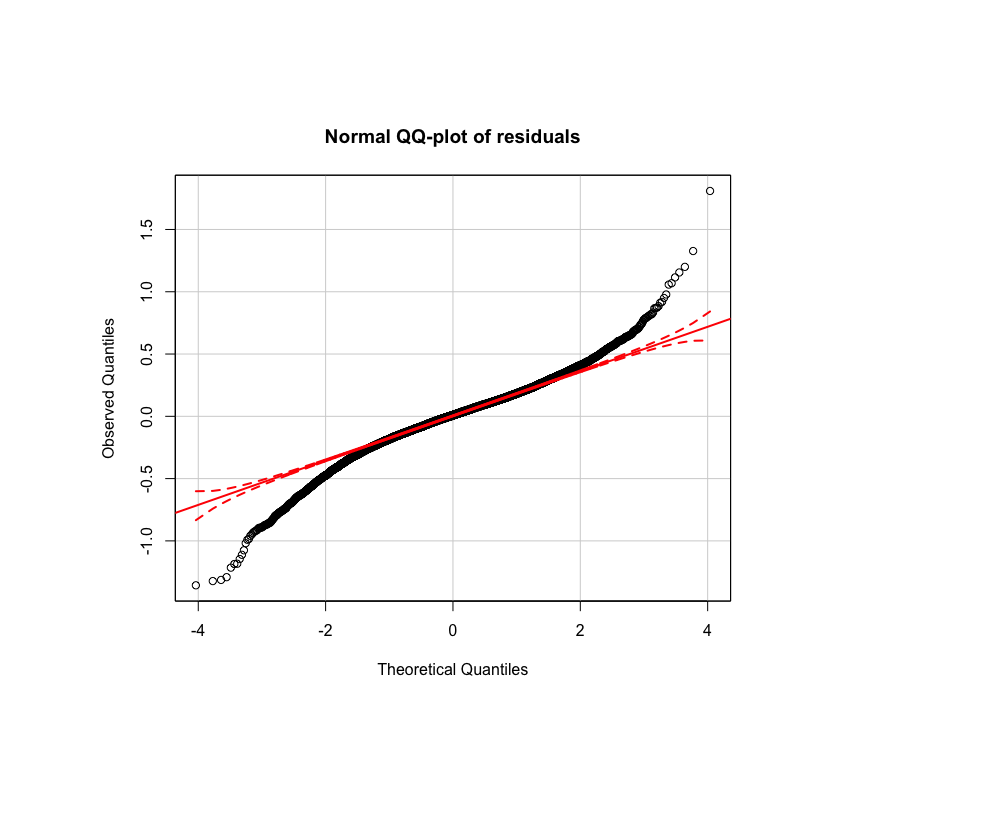}
\caption{A Normal Q--Q plot of the residuals from Implementation 2, with 95\% confidence intervals shown in red.}
\label{fig:joint1_QQ}
\end{center}
\end{figure}

\begin{figure}[H]
\begin{center}
\includegraphics[scale=0.5]{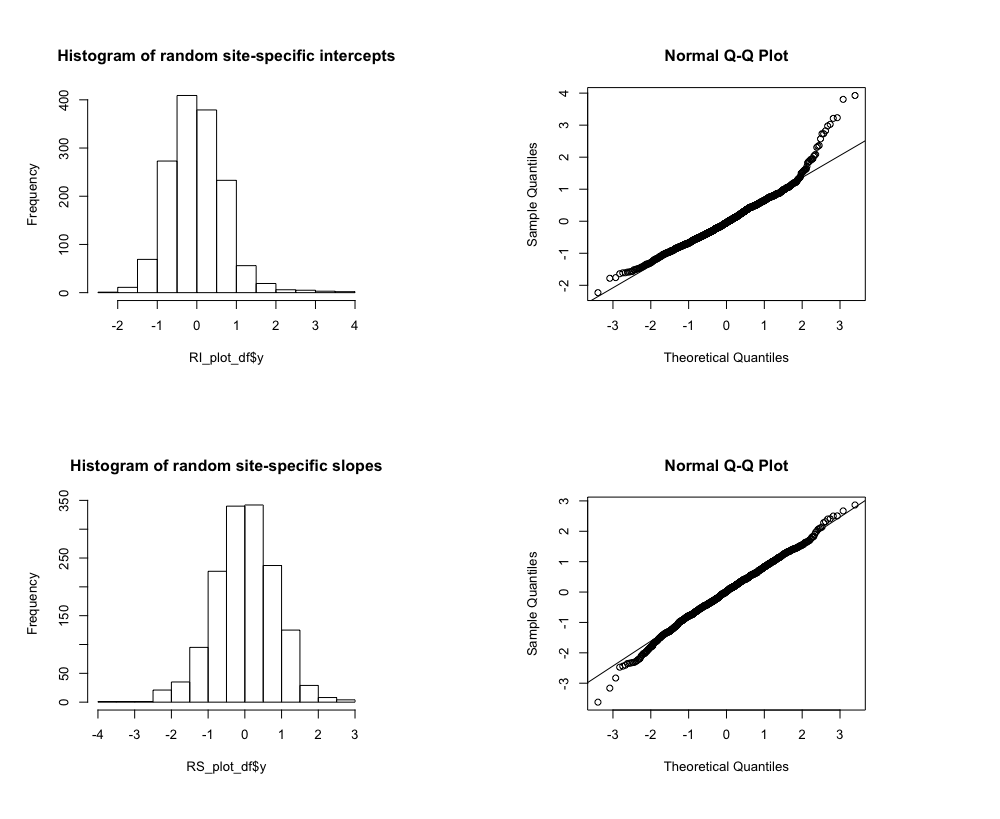}
\caption{Histograms of the spatially--uncorrelated random intercepts (top left) and slopes(bottom left), with corresponding Normal Q--Q plots shown on the right from Implementation 2.}
\label{fig:joint1_RI_plots}
\end{center}
\end{figure}

\begin{figure}[H]
\begin{center}
\includegraphics[scale=0.6]{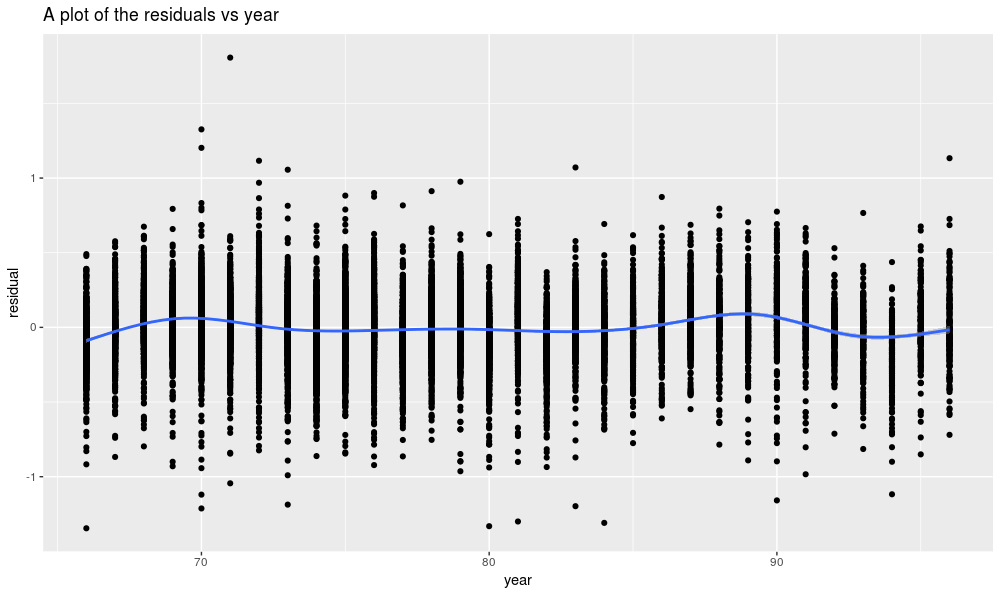}
\caption{A plot of the residuals vs. year for Implementation 3 with a fitted smoother.}
\label{fig:joint2_residuals}
\includegraphics[scale=0.6]{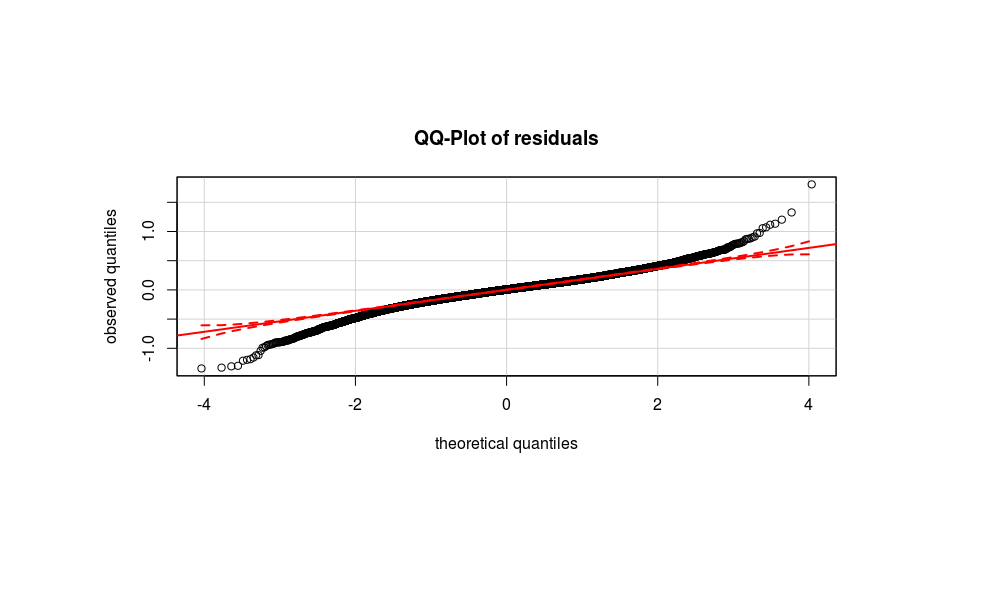}
\caption{A Normal Q-Q plot of the residuals from Implementation 3 with 95\% confidence intervals shown in red.}
\label{fig:joint2_QQ}
\end{center}
\end{figure}

\begin{figure}[H]
\begin{center}
\includegraphics[scale=0.6]{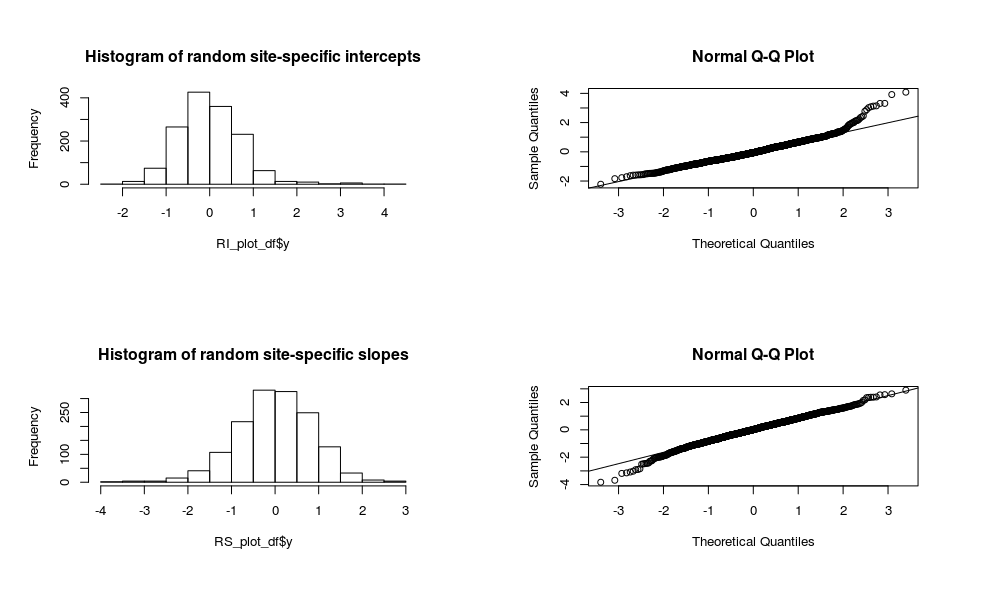}
\caption{Histograms of the spatially-uncorrelated random intercepts (top left) and slopes(bottom right), with corresponding Normal Q-Q plots shown on the right from Implementation 3.}
\label{fig:joint2_RI_plots}
\end{center}
\end{figure}

\end{document}